\documentclass[10pt,sigplan,letterpaper,twocolumn]{article}
\usepackage[10pt,nocopyright]{sigmin}

\usepackage{listings}
\usepackage{color}
\usepackage[hyphens]{url}                                         
\usepackage[breaklinks,colorlinks]{hyperref}                      
\usepackage[usenames,dvipsnames]{xcolor}                          
\hypersetup{citecolor=blue,linkcolor=blue}                        
\usepackage{amsmath,amsopn,amssymb,amsthm}                        
\usepackage{thmtools}
\usepackage{thmbox}
\usepackage[toc,page]{appendix}
\usepackage{subcaption}                                           
\usepackage{endnotes,microtype,xspace,fancyvrb,multirow} 
\usepackage{epigraph} 

\usepackage{booktabs}                                             
\usepackage{array,underscore,relsize}                             
\usepackage[T1]{fontenc}                                          
\usepackage{times}                                                
\usepackage{fancyhdr,lastpage}                                    
\usepackage{enumitem}                                             
\usepackage[labelfont=bf,font=normal,skip=1pt]{caption}           
\usepackage[export]{adjustbox}                                    
\usepackage{breakurl}                                             
\usepackage{pifont}                                               
\usepackage{tablefootnote}                                        
\usepackage{bbding}                                        
\usepackage[linesnumbered,vlined,boxed,commentsnumbered]{algorithm2e}
\DontPrintSemicolon
\newcommand{\ra}[1]{\renewcommand{\arraystretch}{#1}}
\usepackage{amssymb}
\usepackage[normalem]{ulem}
\usepackage[hang,flushmargin]{footmisc}
\usepackage{textcomp}
\usepackage{graphicx}

\usepackage[compact]{titlesec}
\titleformat*{\section}{\large\bfseries}
\titleformat*{\subsection}{\normalsize\bfseries}
\titleformat*{\subsubsection}{\normalsize\bfseries}
\titlespacing*{\section}{0pt}{*2}{*1}
\titlespacing*{\subsection}{0pt}{*1.5}{*0.8}

\hypersetup{
  colorlinks,
  linkcolor={red!50!black},
  citecolor={blue!50!black},
  urlcolor={blue!80!black}
}

\newcommand{\sys}{{\textsc{Mitosis}}}

\newcommand{\legocontainer}{{\textsc{LegoContainer}}}

\newcommand{\fn}{{\textsc{Fn}}}

\newcommand{\fork}{{\mbox{\smaller[0.5]{\textsc{{Fork}}}}}}

\def\ie{i.e.,~}
\def\eg{e.g.,~}

\newcommand{\stitle}[1]{\vspace{1.ex}\noindent{\bf #1}}
\newcommand{\noindentstitle}[1]{\noindent{\bf #1}}
\newcommand{\etitle}[1]{\vspace{0.8ex}\noindent{\em\underline{#1}}}

\newcommand{\blue}[1]{\textcolor{blue}{#1}}

\newcommand{\pcomment}[2]{{\bf[\textcolor{red}{#1}: \textcolor{blue}{#2}]}}

\newcommand{\XD}[1]{\textcolor{blue}{XD: #1}}

\newcommand{\tmac}[1]{{\pcomment{tmac}{#1}}}

\newcommand{\TODO}[1]{\textcolor{red}{TODO: #1}}

\usepackage{tikz}

\usepackage{balance}
\usepackage{authblk}

\begin{document}

\title{\LARGE \bf{No Provisioned Concurrency: Fast RDMA-codesigned Remote Fork\\for Serverless Computing}}

\setlength{\affilsep}{0.5em}
\author[1,2]{Xingda Wei}
\author[1]{Fangming Lu}
\author[1]{Tianxia Wang}
\author[1]{Jinyu Gu}
\author[1]{Yuhan Yang}
\author[1,2]{Rong Chen}
\author[1] {Haibo Chen}
\affil[1]{Institute of Parallel and Distributed Systems, SEIEE, Shanghai Jiao Tong University}
\affil[2]{Shanghai AI Laboratory}



\date{}
\maketitle

\frenchspacing

\begin{abstract}

Serverless platforms essentially face a tradeoff 
between container startup time and provisioned concurrency (\ie{cached instances}),
which is further exaggerated by the frequent need for remote container initialization.  
This paper presents {\sys}, an operating system primitive that provides fast remote fork, 
which exploits a deep codesign of the OS kernel with RDMA. 
By leveraging the fast remote read capability of RDMA 
and partial state transfer across serverless containers, 
{\sys} bridges the performance gap between local and remote container initialization.  
{\sys} is the first to fork over 10,000 new containers from one instance 
across multiple machines within a second,
while allowing the new containers to efficiently transfer the pre-materialized states of the forked one. 
We have implemented {\sys} on Linux and integrated it with {\fn}, 
a popular serverless  platform. 
Under load spikes in real-world serverless workloads, {\sys} reduces the function
tail latency by 89\% with orders of magnitude lower memory usage.
For serverless workflow that requires state transfer, 
{\sys} improves its execution time by 86\%.

\end{abstract}

\section{Introduction}

Serverless computing is an emerging cloud computing paradigm
supported by major cloud providers, including AWS Lambda~\cite{aws-lambda}, 
Azure Functions~\cite{azure-function}, Google Serverless~\cite{google-serverless},
Alibaba Serverless Application Engine~\cite{ali-serverless}
and Huawei Cloud Functions~\cite{huawei-function}.
One of its key promises is \emph{auto-scaling}---users only provide serverless 
functions, and serverless platforms automatically allocate computing resources 
(\eg{containers\footnote{\footnotesize{We focus on executing serverless functions 
with containers in this paper, which is widely adopted 
by existing platforms~\cite{openwhisk,fn,openlambda,DBLP:conf/asplos/JiaW21}.}}}) 
to execute them. 
Auto-scaling makes serverless computing economical:
the platform only bills when functions are executed (no charge for idle time).

\begin{figure}[!t]
	\vspace{-3pt}
	\begin{minipage}{1.\linewidth}
		\includegraphics[scale=1.2]{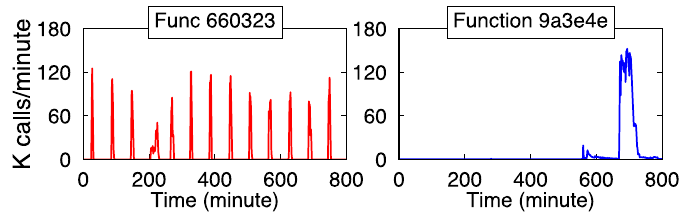} 
	\end{minipage}
	\begin{minipage}{1.\linewidth}
		\hspace{1.5pt}
		\includegraphics[scale=1.2]{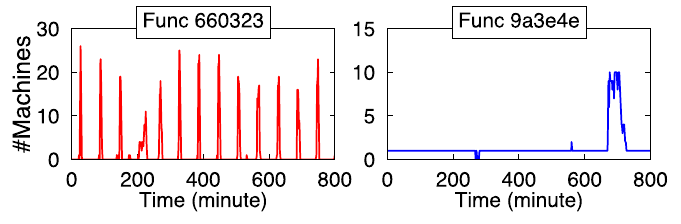} 
	\end{minipage} 
	\begin{minipage}{1\linewidth}
		\vspace{2mm}
	\caption{\small{\emph{The timelines of call frequency (top) and 
	sufficient resource provisioning (bottom) for two serverless functions 
	in a real-world trace from Azure Functions~\cite{DBLP:conf/usenix/ShahradFGCBCLTR20}.
	}}}
	\label{fig:bg-trace}
	\end{minipage} \\[-10pt]
	\end{figure}

However, \emph{coldstart} (\ie{launching a container from scratch for each function}) is 
a key challenge for fast auto-scaling, as the start time (over 100\,ms) can be orders of 
magnitude higher than the execution time 
for ephemeral serverless functions~\cite{du2020catalyzer,sock,216063}.
Accelerating coldstart has become a hot topic in both academia and industry~\cite{AWS-Lambda-Reuse,openwhisk,sock,sand,DBLP:conf/usenix/ShahradFGCBCLTR20,
du2020catalyzer, DBLP:conf/eurosys/AoPV22}. 
Most of them resort to a form of `warmstart' by \emph{provisioned concurrency}, 
\eg{launching a container from a cached one}. 
However, they require non-trivial resources when scaling functions to 
a distributed setting, 
\eg{each machine should deploy many cached containers}.

Unfortunately, scaling functions to multiple machines is common
because a single machine has a limited function capacity to handle 
the timely load spikes. 
Consider the two functions sampled from real-world traces of Azure Functions~\cite{DBLP:conf/usenix/ShahradFGCBCLTR20}. 
The request frequency of function \texttt{9a3e4e} can surge to over 150\,K calls 
per minute, increased by 33,000$\times$ within one minute (see 
the top part of Figure~\ref{fig:bg-trace}).
To avoid stalling numerous newly arriving function calls, 
the platform should immediately launch sufficient containers across 
{multiple machines} (see the bottom part of Figure~\ref{fig:bg-trace}). 
Due to the unpredictable nature of the serverless workload,
it is challenging for the platform to decide the number of cached instances for the warmstart.
Therefore, there is ``no free lunch'' for such resources: 
commercial platforms require users to reserve and pay for them
to achieve better performance (\ie{lower response time}), 
\eg{AWS Lambda Provisioned Concurrency~\cite{aws-pc}}.

Even worse, dependent functions that run in separate 
containers cannot directly transfer states.
Instead, they must resort to message passing or cloud storage for state transfer,
which has data serialization/de-serialization, memory copy and storage stack overheads. 
Recent reports have shown that these may count up to 95\% of the function execution time~\cite{faastlane, DBLP:conf/cidr/HellersteinFGSS19}. 
Unfortunately, transferring states between functions is common 
in serverless workflows---a mechanism to compose functions into more complex applications~\cite{aws-step, openwhisk-composer}.
Though recent research~\cite{faastlane} bypasses such overheads 
for local state transfer (\ie{functions that run on the same machine})
by co-locating local functions in the same container, 
it is still unclear how to do so in a remote setting. 
	
We argue that \emph{remote fork}---forking containers 
across machines like a local fork---is a promising  primitive 
to enable both efficient function launching and fast function state sharing. 
First, the fork mechanism has been shown efficient in both performance and 
resource usage for launching containers on a single machine: 
one cached container is sufficient to start numerous containers 
with 1\,ms time~\cite{sand,du2020catalyzer,DBLP:conf/asplos/DuLJXZC22}. 
By extending the fork mechanism to remote, 
one active container is sufficient to start numerous 
containers efficiently on all the machines.
Second, remote fork provides transparent intermediate state sharing 
between remote functions---the code in the container created by the fork can access 
the pre-materialized states of the forked container transparently
bypassing message passing or cloud storage. 

However, state-of-the-art systems can only achieve a conservative remote fork 
with Checkpoint/Restore techniques (C/R)~\cite{criu,DBLP:conf/asplos/Ustiugov0KBG21}.
Our analysis reveals that they are not  efficient  for serverless computing,
\ie{even slower than coldstart}
due to the costs of checkpointing the memory of parent container into files,
transferring the files through the network 
and accessing the files through a distributed file system (\textsection{\ref{sec:rfork}}).
Even though we have utilized modern interconnects (\ie{RDMA}) to reduce these costs, 
the software overhead of checkpointing and distributed file accesses 
still make C/R underutilize the low latency and high throughput of RDMA.

We present {\sys}, an operating system primitive that provides 
a fast \emph{remote fork} by deeply co-designing with RDMA. 
The key insight is that the OS can directly access the physical memory on remote machines
via RDMA-capable NICs (RNICs)~\cite{tsai2017lite}, which is extremely fast
thanks to bypassing remote OS and remote CPU.
Therefore, we can realize remote fork by imitating local fork
through mapping a child container's virtual memory to 
its parent container's physical memory
without checkpointing the memory.  
The child container can directly
read the parent memory in a copy-on-write fashion using RNIC, 
bypassing the software stacks (\eg{distributed file system}) introduced by traditional C/R.

Leveraging RDMA for remote fork with kernel poses several new challenges (\textsection{\ref{sec:challenges}}):
(1) fast and scalable RDMA-capable connection establishment, 
(2) efficient access control of the parent container's physical memory and 
(3) efficient parent container lifecycle management at scale. 
{\sys} addresses these challenges by 
(1) retrofitting advanced RDMA feature (\ie{DCT~\cite{dct}}),
(2) proposing a new connection-based memory access control method designed for 
remote fork 
and 
(3) co-designing container lifecycle management with the help of serverless platform. 
We also introduce techniques including generalized lean container~\cite{sock}
to reduce containerization overhead for the remote fork.
In summary, 
we show that remote fork can be made efficient, 
feasible and practical on commodity RNICs for serverless computing. 

We implemented {\sys} on Linux with its core functionalities 
written in Rust as a loadable kernel module. 
It can remote-fork $10,000$ containers on 5 machines within 0.86 second.
{\sys} is fully compatible with mainstream containers 
(\eg{runC~\cite{runc}}), making integration with existing container-based 
serverless platforms seamlessly.
To demonstrate the efficiency and efficacy, 
we integrated {\sys} with Fn~\cite{fn}, a popular open-source serverless 
platform. Under load spikes in real-world serverless workloads, 
{\sys} reduces the $99^{th}$ percentile latency of the
spiked function by 89\% with orders of magnitude lower memory usage. 
For a real-world serverless workflow (i.e., FINRA~\cite{finra}) that requires state transfer,
{\sys} reduces its execution time by 86\%. 

\stitle{Contributions}. We highlight the contributions as follows: \\[-18pt]
\begin{itemize}[leftmargin=*,leftmargin=10pt,itemindent=0pt]
    \item \textbf{Problem}: An analysis of
    the performance-resource provisioning trade-off of 
	existing container startup techniques,
    and the costs of state transfer between functions (\textsection{\ref{sec:bg}}).\\[-18pt]
    \item \textbf{{\sys}}: 
    An RDMA-co-designed OS remote fork 
	that quickly launches containers on remote machines without provisioned concurrency
	and enables efficient function state transfer (\textsection{\ref{sec:overview}--\ref{sec:design}}). \\[-18pt]
    \item \textbf{Demonstration}: An implementation on 
	Linux integrated with Fn (\textsection{\ref{sec:serverless}}) 
	and evaluations on both microbenchmarks and real-world serverless applications 
	demonstrate the efficacy of {\sys} (\textsection{\ref{sec:eval}}). 
	{\sys} is publically available at {\small{\url{https://github.com/ProjectMitosisOS}}}. 
\end{itemize}

\section{Background and Motivation}
\label{sec:bg}

\subsection{Serverless computing and container}
\label{sec:bg-serverless}

\begin{table*}[t]  
\centering    
\begin{minipage}{1\linewidth}
\caption{\textit{\small{A comparison of startup techniques for autoscaling 
{$n$} concurrent invocations of one function to {$m$} machines.
\textup{Local} means the resources for the startup are provisioned on the function execution machine. 
The function is a simple python program that prints `hello world'.}}}
\label{tab:motiv-autoscale}
\end{minipage} \\[2pt]  
\ra{1.1}
\centering\small{          
\begin{tabular}{@{~}l | r r r r r@{~}}
\toprule 
& \textbf{Coldstart} & \textbf{Caching} & \textbf{Fork} 
& \textbf{Checkpoint/Restore} & \textbf{Remote fork} \\
& \cite{docker,DBLP:conf/usenix/WangCTWYLDC21} 
& \cite{DBLP:conf/sosp/JiaW21,fn,sock, DBLP:conf/usenix/ShahradFGCBCLTR20,openwhisk}
& \cite{du2020catalyzer,sand,DBLP:conf/asplos/DuLJXZC22}
& \cite{DBLP:conf/eurosys/WangHW19, du2020catalyzer,DBLP:conf/asplos/Ustiugov0KBG21,DBLP:conf/eurosys/AoPV22}
& {\sys} \\
\midrule
\textbf{Local startup performance}  
   & Very slow $(100\,ms)$   & Very fast $(<1\,ms)$
   & Fast $(1\,ms)$          & Medium $(5\,ms)$    & {Fast $(1\,ms)$}  \\  
\textbf{Remote startup performance} 
   & Very slow $(1,000\,ms)$ & N/A 
   & N/A                     & Slow $(24\,ms)$        & {Fast $(3\,ms)$} \\  
\textbf{Overall resource provisioning} 
   & $O(1)$                & $O(n)$               
   & $O(m)$                & $O(1)$                 & {$O(1)$}          \\  
\bottomrule
\end{tabular}
} \\[-5pt]
\end{table*}

Serverless computing is a popular programming paradigm.
It abstracts resource management from the developers:
they only need to write the application as \emph{function}s 
in a popular programming language (\eg{Python}), 
upload these \emph{function}s (as container images) to the platform, 
and specify how to call them. 
The platform can \emph{auto-scale} according to function requests 
by dynamically spawning a container~\cite{openlambda,
fn,ibm-fn,sock,aws-fargate,ali-serverless,azure-function,google-serverless,aws-fargate,
knative}\footnote{\footnotesize{Serverless 
platform may use virtual machines to run functions, which is not the focus of this paper.}} 
to handle each call.
The spawned containers will also be automatically reclaimed after functions return, 
making serverless economical: 
the developers only pay for the in-used containers.  

Container is a popular host for executing functions.
It not only packages the application's dependencies into a single image
that ease the function deployment, 
but also provides lightweight isolation through Linux's \texttt{cgroup}s and 
\texttt{namespace}s, which is necessary to run applications in a multi-tenancy environment.
Unfortunately,
enabling container introduces additional function startup costs and state transferring costs 
due to container bootstrap 
and segregated function address spaces, respectively.


\subsection{Startup and resource provisioning costs}
\label{sec:bg-auto-scaling-costs}

\vspace{-1mm}
\stitle{Coldstart performance cost.}
Starting a container from scratch, commonly named as `\emph{coldstart}',
is notoriously slow. 
The startup includes pulling the container image, 
setting up the container configurations 
and initializing the function language runtime. 
All the above steps are costly, which takes even more than hundreds of milliseconds~\cite{du2020catalyzer,sock}.
As a result, 
coldstart may dominate the end-to-end latency of ephemeral serverless 
functions~\cite{du2020catalyzer,sock,
DBLP:conf/usenix/WangCTWYLDC21,report-state-of-serverless}. For example, 
Lambda@Eedge reports that 67\% of its functions run in less than 
20\,{ms}~\cite{report-state-of-serverless}.
In comparison, starting a Hello-world python 
container with runC~\cite{runc}---a state-of-the-art container runtime---takes 
167\,{ms} and 1783\,{ms} when the container image is stored locally and remotely,
respectively (see Table~\ref{tab:motiv-autoscale}).

\stitle{Warmstart resource cost due to provisioned concurrency.} 
A wealth of researches focus on reducing the startup time of coldstart
with `\emph{warmstart}' techniques~\cite{sock,sand,du2020catalyzer,
DBLP:conf/usenix/ShahradFGCBCLTR20,DBLP:conf/cloud/ThomasAVP20,DBLP:conf/asplos/FuerstS21,
DBLP:conf/usenix/WangCTWYLDC21,DBLP:conf/cloud/YuLDXZLYQ020,FAASM}.
However, they must pay more resource provisioning cost (see Table~\ref{tab:motiv-autoscale}):

\etitle{Caching}~\cite{DBLP:conf/sosp/JiaW21,DBLP:conf/asplos/JiaW21,fn,
AWS-Lambda-Reuse,openwhisk,sock,sand,DBLP:conf/usenix/ShahradFGCBCLTR20}.
By caching finished containers (\eg{via Docker \texttt{pause}~\cite{docker-pause}})
instead of reclaiming them, 
future functions can reuse cached ones (\eg{via Docker \texttt{unpause}}) 
with nearly no startup cost (less than 1\,ms). 
However, 
Caching consumes large in-memory resources: 
the resource provisioned---number of the cached instances ($O(n)$) 
should match the number of concurrent functions ($n$),
because a paused container can only be unpaused once. 
Given the unpredictability of the number of function invocations 
(\eg{load spikes in Figure~\ref{fig:bg-trace}}),
it is challenging for the developers or the platform to decide how 
many cached instances are required. 
Thus,
Caching inevitably faces the trade-off between 
fast startup and low resource provisioning,
resulting in huge cache misses.

\etitle{Fork}~\cite{du2020catalyzer,sand,DBLP:conf/asplos/DuLJXZC22}.
A cached container (\emph{parent}) can call the \texttt{fork} system call (instead of \texttt{unpause})
to start new containers (\emph{children}).
Since fork can be called multiple times, 
each machine only requires one cached instance 
to fork new containers.
Thus, 
fork reduces resource provisioned of Caching---cached containers from $O(n)$ to $O(m)$, 
where $m$ is the number of machines that require function startup.
However,
it is still proportional to the number of machines ($m$)
since fork cannot generalize to a distributed setting. 

\etitle{Checkpoint/Restore (C/R)}~\cite{DBLP:conf/eurosys/WangHW19, du2020catalyzer,
DBLP:conf/asplos/Ustiugov0KBG21}. C/R starts containers from container checkpoints stored 
in a file. It only needs $O(1)$ resource (the file) to warmstart, 
because the file can be transferred through the network if necessary. 
Though being optimal in resource usage, 
C/R is orders of magnitude slower than Caching and fork.
We analyze it in \textsection{\ref{sec:rfork-challenge}} in detail.

\subsection{(Remote) state transfer cost}
\label{sec:bg-state-transfer-costs}

\begin{figure}[!t]
\centering
\vspace{-3.mm}
\includegraphics[scale=1.]{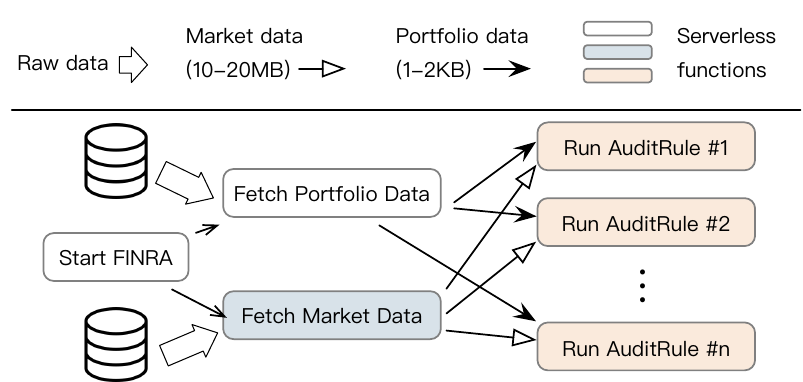} \\[8pt]
\begin{minipage}{1\linewidth}
\caption{\small{\textit{The workflow graph of a real-world serverless application,
Financial Industry Regulatory Authority, FINRA~\cite{finra}.}}}
\label{fig:finra-workflow}
\end{minipage} \\[-15pt]
\end{figure}

\begin{figure*}[!t]
\hspace{1mm}
\includegraphics[scale=1.09,left]{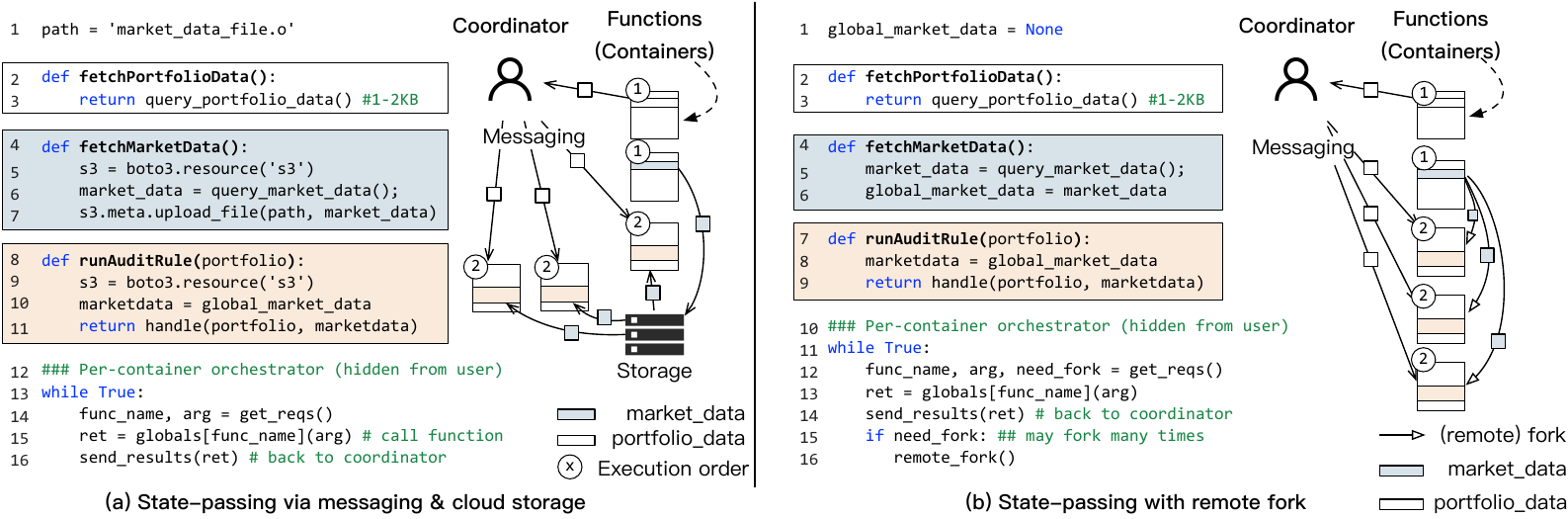} \\[15pt]
\begin{minipage}{1\linewidth}
\caption{\emph{\small{(a) A simplified code of FINRA (see Figure~\ref{fig:finra-workflow}) 
on existing serverless platforms. (b) A simplified code of using (remote) fork to transfer 
states between FINRA functions. \texttt{globals} records a mapping between function 
name and its pointer. 
}}}
\label{fig:finra-fork-code}
\end{minipage} \\[-10pt]
\end{figure*} 

Transferring states between functions is common 
in serverless workflows~\cite{DBLP:conf/asplos/DuLJXZC22, sand, DBLP:conf/middleware/PonsAPSL19,DBLP:conf/asplos/JiaW21, aws-step,openwhisk-composer}. 
A workflow is a graph 
describing the producer-consumer relationships between functions. 
Consider the real-world example FINRA~\cite{finra}
shown in Figure~\ref{fig:finra-workflow}.
It is a financial application that validates trades 
according to the trade (Portfolio) and market (Market) data. 
Upstream functions (the ones that produce states),  
\ie{\texttt{fetchPortfolioData} and \texttt{fetchMarketData}}
first read data from external sources.
Afterward, they transfer the results to 
many downstream functions (the one that consumes states), 
\ie{\texttt{runAuditRule}s} to process them concurrently for a better performance. 

Functions run in different containers can only transfer states either
by copying them through the network via message passing
or exchanging them at a cloud storage service.
Figure~\ref{fig:finra-fork-code} (a) shows a simplified code for 
running FINRA on AWS Lambda.
For small states transfers (less than 32KB, \eg{Portfolio}), 
Lambda piggybacks the states in messages exchanged 
between the coordinator and the function containers~\cite{DBLP:conf/cloud/YuLDXZLYQ020}. 
For large ones (Market), functions must exchange them 
with S3---Lambda's cloud storage service.

Transferring states via messages and cloud storage
inevitably faces the overheads of 
data serialization, memory copies, and cloud storage stacks,
causing up to a 1,000X slowdown~\cite{DBLP:conf/cidr/HellersteinFGSS19,faastlane}. 
To cope with the issue, 
existing work proposes serverless-optimized messaging primitives~\cite{sand} 
or specialized storage systems~\cite{cloudburst, DBLP:journals/usenix-login/KlimovicWSTPK19,DBLP:conf/nsdi/PuVS19},
but none of the mentioned overhead is completely eliminated~\cite{faastlane}.
Faastlane~\cite{faastlane} co-locates functions in the same container with 
\emph{thread}s so that it can bypass these overheads with shared memory accesses. 
However, threads cannot generalize to a distributed setting.
Faastlane fallbacks to message passing if the upstream and downstream
functions are on different machines. 

\section{Remote Fork for Serverless Computing}
\label{sec:rfork}

We show the following two benefits of \emph{remote fork}
to address the issues mentioned in the previous section.

\etitle{Efficient (remote) function launching.}
When generalizing the {\fork} primitive to a remote setting, a single \emph{parent}
container is sufficient to launch subsequent \emph{child}\footnote{\footnotesize{
    We may also call the kernel/machine hosting the parent/child container as \emph{parent}/\emph{child}
    in this paper without losing generality.}}  containers across the cluster, 
similar to C/R (see Table~\ref{tab:motiv-autoscale}). 
We believe $O(1)$ resource provisioning is desirable for the developers/tenants 
since they only need to specify whether they need resource for warmstart, 
instead of how many (\eg{the number of machines for forking or cached instances~\cite{aws-pc} 
for Caching}).

\etitle{Fast and transparent (remote) state transfer.}
The {\fork} primitive essentially bridges the address spaces of parent 
and child containers. 
The transferred states are pre-materialized in the parent memory, 
so the child can seamlessly access them with shared memory abstraction 
with no data serialization,
zero-copy (for read-only accesses\footnote{\footnotesize{In the 
case of the traditional fork. 
{\sys} further optimizes with one-sided RDMA (\textsection{\ref{sec:overview}}),
 allowing zero-copy even for read-write accesses. }})
and cloud storage costs. 
Meanwhile, the \emph{copy-on-write} semantic in the {\fork} primitive
avoids the costly memory coherence protocol in traditional distributed 
shared memory systems~\cite{DBLP:journals/tocs/LiH89,DBLP:journals/jcst/HongZYZGC19}. 

Figure~\ref{fig:finra-fork-code} (b) presents a concrete example of using fork 
to transfer the market data in FINRA (see Figure~\ref{fig:finra-workflow}).
Suppose all functions are packaged in the same container\footnote{\footnotesize{Commonly 
found in serverless platforms~\cite{knative,faastlane,openwhisk-composer}. }}, and 
the container has an orchestrator dispatching function requests to 
user-implemented functions (lines 11--14). 
We further assume the coordinator issuing requests 
to the orchestrators is fork-aware (\textsection{\ref{sec:fork-aware}}):
based on the function dependencies in the workflow graph (e.g., Figure~\ref{fig:finra-workflow}),
it will request the orchestrator to fork children if necessary (line 12).
After the orchestrator finishes \texttt{fetchMarketData} (line 13), 
it forks (lines 15--16) to run downstream functions (\texttt{runAuditRule}), 
which can directly access the \texttt{global\_market\_data} pre-materialized by the parent (line 8).

\begin{figure}[t]
\includegraphics[scale=1.0,center]{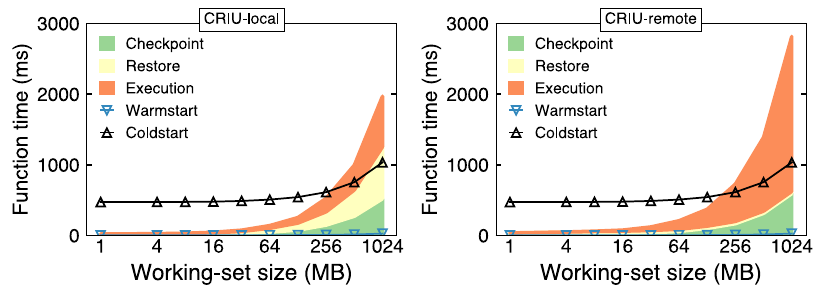} \\[5pt]
\begin{minipage}{1\linewidth}
\caption{\small{\emph{Analysis of using C/R for remote fork. \textbf{\textup{Setup}}: 
{CRIU-local}: CRIU with a local file system (e.g., tmpfs), which uses
RDMA to transfer files between machines.
{CRIU-remote}: CRIU with an RDMA-accelerated distributed file system 
(e.g., Ceph~\cite{ceph-rdma})}.
}}
\label{fig:motiv-criu}
\end{minipage} \\[-15pt]
\end{figure}

\begin{figure*}[!t]     
\centering\includegraphics[scale=1.]{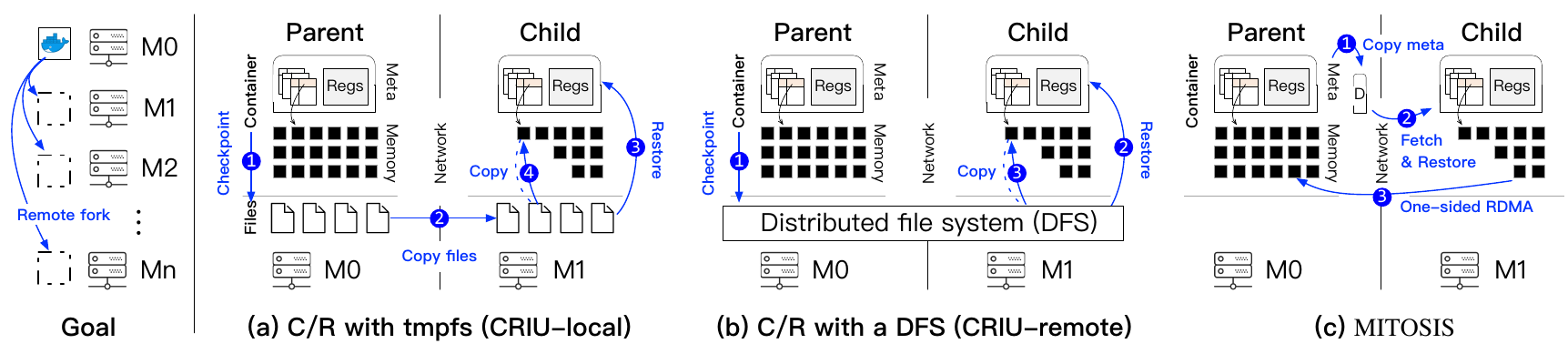}  \\[5pt]
\begin{minipage}{1\linewidth}
\caption{\small{\emph{An overview of different approaches to achieve 
ultra-fast remote fork, including (a) C/R with a local filesystem (\eg{tmpfs}), 
(b) C/R with a fast distributed filesystem (\eg{Ceph~\cite{ceph}}), and
(c) {\sys}. }}}
\label{fig:overview} 
\end{minipage} \\[-10pt]
\end{figure*}

\label{sec:rfork-challenge}
\stitle{Challenge: remote fork efficiency.}
To the best of our knowledge, existing containers can only remote fork
with a C/R-based approach~\cite{smith1987implementing,criu-rfork}. 
To fork a child, the parent first \emph{checkpoints} its states 
(\eg{register values and memory pages}) by copying them to a file,
and then \emph{transfers} the file to the child---either using 
a remote file copy---see \emph{CRIU-local} in Figure~\ref{fig:overview} (a), or 
a distributed file system (see \emph{CRIU-remote} in Figure~\ref{fig:overview} (b)). 
After receiving the file, 
the child \emph{restores} the parent's execution by loading the container states
from the checkpointed file. 
Note that C/R may load some states (\ie{memory pages}) on-demand 
for better performance~\cite{DBLP:conf/eurosys/WangHW19}. 

Unfortunately, the C/R-based remote fork is not efficient enough for serverless computing. 
Figure~\ref{fig:motiv-criu} (a) shows the execution time of serverless functions 
on a remote machine using CRIU~\cite{criu}---the state-of-the-art C/R 
on Linux (with careful optimizations, see \textsection{\ref{sec:eval}} for details)
to realize CRIU-local and CRIU-remote.
The synthetic function randomly touches the entire parent's memory.
We observe that remote fork can even be 2.7X slower than coldstart 
if it accesses 1\,GB remote memory.
We attribute it to one or more of the following issues. 

\etitle{Checkpoint container memory.}
CRIU takes 9\,ms (resp. 518\,{ms}) and 15.5\,{ms} (resp. 590\,{ms}) to checkpoint 
1\,{MB} (resp. 1\,{GB}) memory of the parent container using local or distributed 
file systems, respectively. 
The overhead is dominated by copying the memory to the files:
unlike the local fork, the child's OS resides on another machine 
and thus, lacks direct memory access capability to the parent's memory pages.

\etitle{Copy checkpointed file.}
For CRIU-local, transferring the entire file from the parent to the child 
takes 11--734\,{ms} for 1\,{MB}--1\,{GB} image (compared to the 0.61--570\,{ms} 
execution time), respectively. 
The whole file copy is typically unnecessary 
since serverless functions typically access a partial state of the parent container~\cite{DBLP:conf/eurosys/WangHW19} 
(see also Figure~\ref{fig:eval-cow} (b)).

\etitle{Additional restore software overhead.}
CRIU-remote enables on-demand file transfer\footnote{\footnotesize{CRIU lazy 
migration~\cite{criu-lazy} also supports on-demand transfer. However, 
it is not optimized for RDMA and is orders of magnitude slower than our evaluated CRIU-remote 
(210 vs. 42\,ms) for the python hello function.}}:
it only reads the required remote file pages during page faults.
However, the execution time is 1.3--3.1$\times$ longer than CRIU-local 
because each page fault requires a DFS request to read the page:
the DFS latency (100\,$\mu$s) is much higher than local file accesses.
More importantly, the latency is much higher than one network round-trip time
(3\,$\mu$s) due to the software overhead.

\section{The {\sys} Operating System Primitive}
\label{sec:overview}


\stitle{Opportunity: kernel-space RDMA.} 
Remote Direct Memory Access (RDMA) is a fast networking feature 
widely deployed in data-centers~\cite{tsai2017lite,DBLP:conf/sigcomm/GuoWDSYPL16,DBLP:conf/nsdi/GaoLTXZPLWLYFZL21}. 
Though commonly used in the user-space,
RDMA further gives the kernel the ability to read/write the \emph{physical memory} of remote machines~\cite{tsai2017lite}  
bypassing remote CPUs (\ie{one-sided RDMA READ}),
with low latency (\eg{2\,$\mu$s}) and high bandwidth (400\,Gbps). 

\stitle{Approach: imitate fork with RDMA.}
{\sys} achieves an efficient remote fork by imitating the local fork with RDMA.
Figure~\ref{fig:overview} (c) shows an overview.
First, we copy the parent’s metadata (e.g., page table) 
to a condensed descriptor (\textsection{\ref{sec:prepare}}) to fork a child ({\blue{\ding{182}}}).
Note that unlike C/R, we don't copy the parent's memory pages to the descriptor. 
The descriptor is then copied to the child via RDMA
to recover the parent's metadata,
similar to \texttt{copy_process} in the local fork ({\blue{\ding{183}}}). 
During execution, we configure the child's remote memory accesses to trigger page faults, 
and the kernel will read the remote pages accordingly.
The fault handler is triggered naturally in an on-demand pattern,
which avoids transferring the entire container state.
Meanwhile, 
{\sys} directly uses one-sided RDMA READ to read the remote physical memory ({\blue{\ding{184}}}), 
bypassing all the software overheads.

\begin{figure}[!t]
\vspace{-5mm}
\hspace{-2mm}
\includegraphics[scale=1.6]{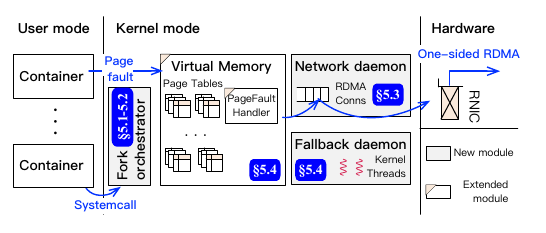} \\[-5pt]
\begin{minipage}{1\linewidth}
\caption{\small{\textit{The {\sys} architecture.}}}
\label{fig:arch}        
\end{minipage} \\[-25pt]
\end{figure}

\stitle{Architecture.} 
We target a decentralized architecture---each machine can fork from others and vice versa.
Note that we do not require dedicated resources (\eg{pinned memory}) to fork containers,
thus, non-serverless applications can co-run with {\sys}. 
We realize {\sys} by adding four components to the kernel (see Figure~\ref{fig:arch}): 
The \emph{fork orchestrator} rehearsals the remote fork execution (\textsection{\ref{sec:prepare} and \ref{sec:resume}}).
The \emph{network daemon} manages a scalable RDMA connection pool (\textsection{\ref{sec:network}})
for communicating between kernels.
We extend OS's 
\emph{virtual memory subsystems} to utilize the remote memory with RDMA (\textsection{\ref{sec:vma}}).
Finally, 
\emph{fallback daemon} provides RPC handlers to restore rare remote memory accesses 
that cannot utilize RDMA. 

\stitle{Security model}. 
We preserve the security model of containers,
\ie{the OS and hardware (RNIC) are trustworthy while malicious containers (functions) may exist}.

\subsection{Challenges and approaches}
\label{sec:challenges}

\stitle{Efficient and scalable RDMA connection setup.}
Though RDMA is fast (\eg{2\,$\mu$s}), 
it is traditionally only supported in the connection-oriented transport (RC),
where connection establishment is much slower (\eg{4\,ms}~\cite{krcore} 
with a limited 700 connections/second throughput).
Caching connections to other machines can mitigate the issue, 
but it is impractical
when RDMA-capable clusters have scaled to more than 10,000 nodes~\cite{DBLP:conf/nsdi/GaoLTXZPLWLYFZL21}.

We retrofit DCT~\cite{dct}, 
an underutilized but widely supported advanced RDMA feature
with fast and scalable connection setups
to carry out communications between kernels
(\textsection{\ref{sec:network}}). 

\stitle{Efficient remote physical memory control.}
{\sys} exposes the parent's physical memory to the children
for the fastest remote fork. 
However, this approach introduces consistency problems in corner cases. 
If the OS changes a parent's 
virtual--physical mappings~\cite{ksm, thp, swap, page-migrate} (\eg{swap}~\cite{swap}),
the children will read an incorrect page. 
User-space RDMA can use memory registration (MR)~\cite{rdma-manual} for the access control.
However, MR has non-trivial registration overheads~\cite{DBLP:conf/asplos/GuoSLHZ22}.
Further, kernel-space RDMA has limited support for MR---it
only supports MR on RCQP (with FRMR~\cite{kernel-verbs}).

We propose a registration-free memory control method (\textsection{\ref{sec:vma}}) that 
transforms RNIC's memory checks to connection permission checks.
We further make the checks efficient by utilizing
DCT's scalable connection setup feature. 

\stitle{Parent container lifecycle management.}
For correctness, 
we must ensure a forked container (parent) is alive 
until all its successors (including children forked from the children) finish. 
A naive approach is letting each machine track the lifecycles of 
the successors of its hosting parents. 
However, it would pose significant management burdens: 
a parent's successors may span multiple machines, 
forming a distributed \emph{fork tree}. 
Meanwhile, each machine may have multiple trees. 
Consequently, 
each machine needs extensive communications with the others 
following paths in the trees to ensure a parent can be safely reclaimed. 

To this end,
we onload the lifecycle management to the serverless platform (\textsection{\ref{sec:fork-tree}}).
The observation is that serverless coordinators (nodes
that invoke functions via fork) naturally 
maintain the runtime information of the forked containers.
Thus, they can trivially decide when to reclaim parents.

\section{Design and Implementation} 
\label{sec:design}


For simplicity, we first assume one-hop fork (\ie{no cascading})
and then extend to multi-hops fork (see \textsection{\ref{sec:multi-fork}}).

\begin{figure}[!t]
        \vspace{0.5mm}
        \hspace{-1.0mm}
        \includegraphics[scale=0.89]{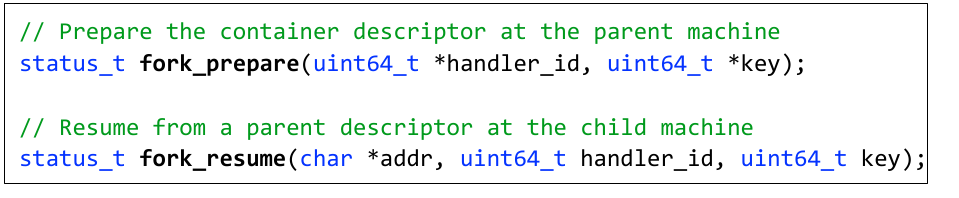} \\[2pt]
        \begin{minipage}{1\linewidth}
        \caption{\small\emph{The major {\sys} remote fork system calls.
        }}
        \label{fig:api}
        \end{minipage} \\[-25pt]
        \end{figure}

\stitle{API.}
We decouple the fork into two phases (see Figure~\ref{fig:api}):
The user can first call \texttt{fork\_prepare} 
to generate the parent's metadata (called \emph{descriptor}) related to remote fork.
The descriptor is globally identified by the local unique \texttt{handle\_id} and \texttt{key}
(generated and returned by the prepared call) and the parent machine's RDMA address.
Given the identifier, 
users can start a child via \texttt{fork\_resume} 
at another machine (can be the same as the parent, \ie{local fork}). 

Compared to the traditional one-stage fork system call,
a two-phase fork API (prepare and resume)---similar to \texttt{pause} and \texttt{unpause} in Caching
is more flexible for serverless computing.
For example, {after preparing and recording the parent's identifier at the coordinator, 
it can later start children without communicating with the parent machine}. 

\stitle{Visibility of the parent's data structures.} 
By default, {\sys} exposes all the parent's data structures---including 
virtual memory and file descriptors, to the child after \texttt{fork\_prepare}. 
{\sys} could introduce APIs to let the application limit the scope of the exposure, 
but currently, we find it unnecessary: 
parents must trust the children since they are from the same application. 

\subsection{Fork prepare}
\label{sec:prepare}

\texttt{fork\_prepare} will generate a local in-memory data structure (\emph{container descriptor})
capturing the parent states, which contains
(1) cgroup configurations and namespace flags---for containerization, 
(2) CPU register values---for recovering the execution states, 
(3) page table and virtual memory areas (VMAs)---for restoring the virtual memory,
and (4) opened file information---for recovering the I/O.
We follow local fork (\eg{Linux's \texttt{copy_process()})
to capture (1)--(3) and CRIU~\cite{criu} for (4). 
Since deciding when to reclaim a descriptor is challenging,  
we always keep the prepared parents (and their descriptors) alive
unless the serverless platform explicitly frees them (\ie{via \texttt{fork\_reclaim}}).

Though the descriptor plays a similar role as C/R checkpointed file, 
we emphasize one key difference: 
the descriptor only stores the page table, not the memory pages. 
As a result, 
it is orders of magnitude smaller (KB vs. MB)
and orders of magnitude faster to generate and transfer.

\subsection{Fork resume}
\label{sec:resume}

\texttt{fork\_resume} resumes the parent's execution state 
by fetching the parent descriptor and then restoring from it.
We now describe how to make the above two steps fast. 
For now, we assume the child OS has established network connections 
capable of issuing RPCs and one-sided RDMAs to the parent. 
The next section describes the connection setup. 

\stitle{Fast descriptor fetch with one-sided RDMA. }
A straightforward implementation of fetching the descriptor 
is using RPC.
However, RPC's memory copy overhead is non-trivial (see Figure~\ref{fig:e2e-factor-analysis}),
as the descriptor of a moderate-sized container may consume several KBs. 
The ideal fetch is using one one-sided RDMA READ, which requires 
(1) storing the parent's descriptor into a consecutive memory area and 
(2) informing the child's OS of the memory's address and size in advance. 

The first requirement can be trivially achieved by 
serializing the descriptor into a well-format message. 
Data serialization has little cost (sub-millisecond) due to the simple data structure of descriptor.  
For the second requirement, 
a naive solution is to 
encode the memory information in the descriptor identifier (\eg{\texttt{handler\_id}})
that is directly passed to the resume system call.
However, this approach is insecure 
because a malicious user could pass a malformed ID, 
causing the child to read and use a malformed descriptor. 
We adopt a simple solution to remedy this: 
{\sys} will send an authentication RPC 
to query the descriptor memory information with the descriptor identifier. 
If the authentication passes, 
the parent will send back the descriptor's stored address and payload 
so that the child can directly read it with one-sided RDMA. 
We chose a simple design because
the overhead of an additional RPC (several bytes) is typically negligible:
reading the descriptor (several KBs) will dominate the fetch time.

\stitle{Fast restore with generalized lean containers.}
With the fetched descriptor, 
child OS uses the following two steps to resume a child to the parent's execution states:
(1) Containerization: set the \texttt{cgroups} and \texttt{namespaces} to match the parent's setup; 
(2) Switch: replace the caller's CPU registers, page table, and I/O descriptors with the parent's.
The switch is efficient (finishes in sub-milliseconds): 
it just imitates the local fork---\eg{unmapping the caller's current memory mapping 
and mapping the child's virtual memory 
to the parents by copying parent's page table to the child} .
On the other hand, containerization can take tens of milliseconds 
due to the cost of setting \texttt{cgroup}s and \texttt{namespace}s. 

Fortunately, fast containerization has been well-studied~\cite{sock,sand,DBLP:conf/eurosys/CaddenUADKA20,216057}.
For instance, 
SOCK~\cite{sock} introduces \emph{lean container},
which is a special container having the minimal 
configurations necessary for serverless computing. 
It further uses pooling to hide the cost of container bootstrap,
reducing its time from tens of milliseconds to a few milliseconds.
We generalize SOCK's lean container to a distributed setting to
accelerate the containerization of the remote fork.
Specifically, before resuming a remote parent, 
we will use SOCK to create an empty lean container 
that satisfies the parent's isolation requirements. 
Afterward, the empty container calls {\sys} to resume execution. 
Since the container has been properly configured with SOCK, 
we can skip the costly containerization. 

\subsection{Network daemon}
\label{sec:network}

\begin{figure}[!t]
    \centering
    \includegraphics[scale=1.15]{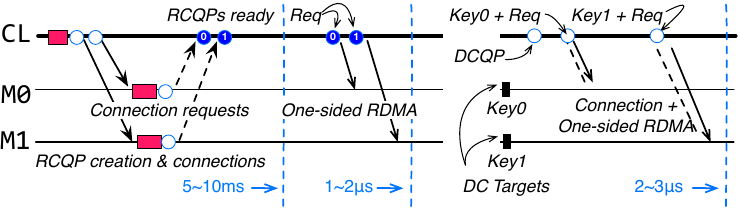} \\[8pt]
    \begin{minipage}{1\linewidth}
    \caption{\small\emph{A comparison of a client (CL) using two RCQPs and 
    DCQP to communicate with two machines (M1 and M2).
    }}
    \label{fig:dct}
    \end{minipage} \\[-15pt]
    \end{figure}

The network daemon aims to reduce the costs of creating RDMA connections
(commonly called \emph{RCQP}) 
on the critical path of the remote fork. 
Meanwhile, it also avoids caching RCQPs connected to all the servers
to save memory.

\stitle{Solution: Retrofit advanced RDMA transport (DCT).}
The essential requirement behind the goal is that 
we need QP to be connectionless. 
RDMA does provide a connectionless transport---unreliable datagram (UD),
but it only supports messaging, so we can just use it for RPC.

We find dynamic connected transport (DCT)~\cite{dct}---a less studied but widely supported 
RDMA feature suits remote fork well. 
DCT preserves the functionality of RC 
and further provides a connectionless illusion:
a single DCQP can communicate with different nodes. 
The target node only needs to create a \emph{DC target}, 
which is identified by the node's RDMA address and a 12B \emph{DC key}\footnote{\footnotesize{Consist 
of a 4B NIC-generated number and 8B user-passed key. }}. 
After knowing the keys, a child node can send one-sided RDMA requests to the corresponding targets
without connection---the hardware will piggyback the connection with 
data processing and is extremely fast (within 1$\mu$s~\cite{krcore,fasst}),
as shown in Figure~\ref{fig:dct}.

Based on DCT, the network daemon manages a small kernel-space DCQP pool for 
handling RDMA requests from children. 
Typically, one DCQP per-CPU is sufficient to utilize RDMA~\cite{krcore}.
However, 
using DCT alone is insufficient because 
the child needs to know the DCT key in advance to communicate with the parent.
Therefore, we also implement a kernel-space FaSST RPC~\cite{fasst} 
to bootstrap DCT.
FaSST is a UD-based RPC that supports connectionless. 
With RPC, 
we piggyback the DCT key associated with the parent in the 
RPC request to query the parent's descriptor. 
To save CPU resources, we only deploy two kernel threads 
to handle RPCs, which is sufficient for our workloads (see Figure~\ref{fig:eval-e2e-thpt} (b)).

\stitle{Discussion on DCT overheads. } 
DCT has known performance issue due to extra reconnection messages.
Compared with RC, it causes up to 
55.3\% performance degradations for small (32B) one-sided RDMA READs~\cite{fasst}. 
Nevertheless, the reconnection has no effect on the large (\eg{more than 1\,KB}) transfer 
because transferring data dominates the time~\cite{krcore}. 
Since the workload pattern of {\sys} is dominated by large transfers, 
\eg{reading remote pages in 4KB granularity}, 
we empirically found no influence from this issue.  

\subsection{RDMA-Aware virtual memory management}
\label{sec:vma}

For resume efficiency, we directly set the page table entries (PTE)
of the children's mapped pages to the parent's physical addresses (PA)
during the resume phase.
However, the original OS is unaware of the remote PA in the PTE.
Thus, 
we dedicate a remote bit in the PTE for distinction. 
In particular,
the OS will set the remote bit to be 1 and clear the present bit of the PTE
during the switch process at the resume phase. 
Afterward,
child's remote page access will trap in the kernel after the switch. 
Consequently, 
{\sys} can handle them in the RDMA-aware page fault handler.
Note that we don't change the table entry data structure:
we utilize an ignored PTE bit (\ie{one in $[58:52]$~\cite{intel-isa}}) for the remote bit. 

\begin{table}[t]  
\centering    
\vspace{2mm}  
\begin{minipage}{1\linewidth}
\caption{{\small\emph{
A summary of page fault handling related to remote fork at the child 
categorized by whether the virtual address (VA) is mapped to remote 
and whether the remote physical address (PA) is stored.
}}}
\label{tab:restore-policy}
\end{minipage} \\[2pt]          
\ra{1.1}
\centering\small{            
\resizebox{.45\textwidth}{!}{%
\begin{tabular}{@{~}l | c c r@{~}}
\toprule
\textbf{Example}  & \textbf{VA mapped} & \textbf{Parent PA in PTE} & \textbf{Method} \\ 
\midrule
\textbf{Stack grows}       & {No}   & {No}      & Local         \\ 
\textbf{Code in .text}{~}  & {Yes}  & {Yes}     & RDMA          \\ 
\textbf{Mapped file}       & {Yes}  & {No}      & RPC           \\
\bottomrule
\end{tabular}%
}
} \\[-5pt]
\end{table}

\stitle{RDMA-aware page fault handler.}
Table~\ref{tab:restore-policy} summarizes how we handle different faults
related to remote fork. 
If the fault page has not mapped to the parent, \eg{stack grows}, 
we handle it locally like a normal page fault. 
Otherwise, we check whether the fault virtual address (VA) 
has a mapped remote PA. 
If so, we use one-sided RDMA to read the remote page to a local page. 
Most child pages can be restored via RDMA because 
serverless function typically touches a subset of the previous run~\cite{DBLP:conf/eurosys/WangHW19, du2020catalyzer}. 
In case of a missed mapping, we fallback to RPC. 

\stitle{Fallback daemon.}
Each node hosts a fallback daemon that spawns kernel threads 
to handle children's paging requests,
which contains the parent identifier and the requested virtual address. 
The fallback logic is simple:
After checking the validity of the request, 
the daemon thread will load the page on behalf of the parent. 
If the load succeeds, 
we will send the result back to the child.

\stitle{Connection-based memory access control and isolation.} 
Direct exposing the parent's physical memory improves the remote fork speed. 
Nevertheless, we need to reject accesses to mapped pages that no longer belong to a parent
and properly isolate accesses to different containers. 
Since we expose the memory via one-sided RDMA in a CPU-bypassing way,
we can only leverage RNIC for the control. 

{\sys} proposes a connection-based memory access control method. 
Specifically, 
we assign different RDMA connections to different portions of 
the parent's virtual memory area (VMA), \eg{one connection per VMA}. 
If a mapped physical page no longer belongs to a parent, 
we will destroy the connection related to the page's VMA.
Consequently, the child's access to the page will be rejected by the RNIC. 
The connections are all managed in the kernel 
to prevent malicious users from accessing the wrong remote container memory. 

To make connection-based access control practical, 
each connection must be efficient in creation and storage. 
Fortunately, the DCQP satisfies these requirements well.  
At the child-side, each connection (DC key) only consumes 12B---different DC connections
can share the same DCQP.
Meanwhile, the parent-side DC target consumes 144B. 
Note that creating DCQPs and targets also has overheads. 
Yet, they are logically independent of the parent's memory. 
Therefore, we use pooling to amortize their creation time (several ms). 

\begin{figure}[!t]
    \vspace{-2mm} 
            \hspace{-3.2mm}
            \includegraphics[scale=1.5]{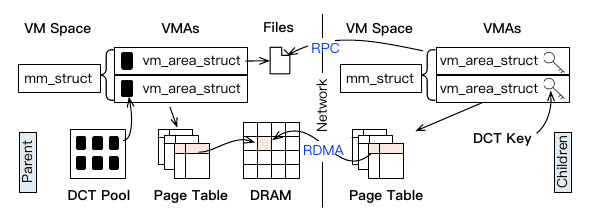} \\[3pt]
            \begin{minipage}{1\linewidth}
            \caption{\small\emph{An illustration of the extended virtual memory subsystems
            to map children's virtual addresses to remote memory. 
            The memory space is divided into a list of virtual memory area (VMA)s, each managed by a \texttt{vm\_area\_struct}. 
            DC target pool and DCT keys are used by connection-based memory access control. 
            }}
            \label{fig:vma}
            \end{minipage} \\[-10pt]
            \end{figure}

Figure~\ref{fig:vma} shows the DCT-based access control in action.
Upon fork preparation, {\sys} assigns one DC target---selected from a target pool---to each parent VMA.
The pool is initialized during boot time and is periodically filled in the background. 
The DC keys of these targets are piggybacked in the parent's descriptor 
so that the children can record them in their VMA during resume. 
Upon reading a parent's page, the child will use the key corresponding to the page's VMA 
to issue the RDMA request. 
With this scheme, 
if the parent wants to reject accesses to this page, 
it can destroy the corresponding DC target. 

Connection-based control has false positives:
after destroying a VMA's assigned target, all remote accesses to it are rejected. 
Assigning DC targets in a more fine-grained way (\eg{multiple targets per VMA})
can mitigate the issue at the cost of increased memory usage. 
We found it is unnecessary 
because VA--PA changes are rare at the parent. 
For example, swap never happens if the OS has sufficient memory. 

\stitle{Security analysis.}
Compared with normal containers, 
{\sys} additionally exposes its physical memory to remote machines via RDMA. 
Nevertheless, since remote containers must leverage their kernels to read the exposed memory,  
a malicious container cannot read others states as long as its kernel is not compromised. 
Besides this, the inherent security issues of RDMA~\cite{SRDMA,ReDMArk,Bedrock} may also endanger {\sys}.
While such security threats are out of the scope of our work, 
it is possible to integrate orthogonal solutions~\cite{SRDMA,ReDMArk,Bedrock,tsai2017lite} 
to improve the security of {\sys}.


\stitle{Optimizations: prefetching and caching.} 
Even with RDMA, reading remote pages is still much slower than local memory accesses~\cite{farm} 
(3\,$\mu$s vs. 100\,ns). 
Thus, we apply two standard optimizations: 
\emph{Prefetching} prefetches adjacent remote pages upon page faults. 
Empirically, we found a prefetch size of one is sufficient to improve 
the performance of remote fork at a small cost to the runtime memory (see Figure~\ref{fig:eval-prefetch}). 
Thus, {\sys} only prefetches one adjacent page by default. 
\emph{Caching} caches the finished children's page table (and the read pages) 
in the kernel. 
A later child forking the same parent can then reuse the page table 
in a copy-on-write way to avoid reading the touched pages again.
This is essentially a combination of local-remote fork. 
To avoid extra memory cost, we only keep the cached page table for a short period 
(usually several seconds) 
to cope with load spikes (\eg{see Figure~\ref{fig:bg-trace}}). 

\subsection{Supporting multi-hops remote fork}
\label{sec:multi-fork}

\begin{figure}[!t]
        \hspace{0mm}
        \includegraphics[scale=1.1]{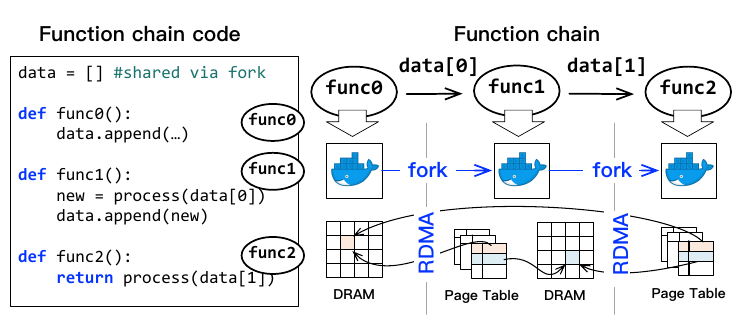} \\[1pt]
        \begin{minipage}{1\linewidth}
        \caption{\small\emph{An illustration of multi-hops remote fork. 
        }}
        \label{fig:mfork}
        \end{minipage} \\[-25pt]
        \end{figure}

{\sys} supports multi-hops fork:
a child can be forked again with its children possibly on the third machine. 
It is similar to one-hop fork except that 
we need to further track 
the ownership of remote pages in a fine-grained way.  
As shown in Figure~\ref{fig:mfork}, 
the pages behind \texttt{data[1]} and \texttt{data[0]} 
resides on two different machines. 
A naive approach would be maintaining a map to track the owner of each virtual page. 
However, it would consume non-trivial storage overhead.
To reduce memory usage, 
{\sys} encodes the owner in the PTE: 
we dedicate 4\,bits in the PTE's ignored bits 
to encode the remote page machine---supporting a maximum of 15-hops remote fork
(up to 15 ancestors)

\section{Bringing {\sys} to Serverless Computing}
\label{sec:serverless}

This section describes how we apply {\sys} to 
{\fn}~\cite{fn}---a popular open source serverless platform.
Though we focus on {\fn},
we believe our methodology can also apply to other serverless platforms (\eg{OpenWhisk~\cite{openwhisk}})
because they follow a similar system architecture (see Figure~\ref{fig:fn}). 

\stitle{Basic {\fn}.}
Figure~\ref{fig:fn} shows an overview of {\fn}. 
It handles the function request that is either an invocation of a single function,
or an execution of a serverless workflow (\eg{see Figure~\ref{fig:finra-workflow}}).
A dedicated \emph{coordinator} is responsible for scheduling the 
executions of these requests. 
The function code must be packed to a container and 
uploaded to a Docker registry~\cite{docker-registry} managed by the platform.  

To handle the invocation of a single function, 
the coordinator will direct the request to an \emph{invoker} 
chosen from a pool of servers. 
After receiving the request, 
the invoker spawns a container with Caching to accelerate startups 
to execute the function.
Note that {\fn} hides the mapping of request to user-function
(\eg{12--16 in Figure~\ref{fig:finra-fork-code}} (a))
with function development kit (\emph{FDK}):
\ie{the user only needs to provide the code for the function,
not the code that dispatches the requests to the function. 
Thanks to this abstraction,
we can extend FDK to add the fork capabilities. 

To execute a workflow, 
the coordinator will first decompose the workflow into
single-function calls (one for each workflow graph node),
then schedule them based on the dependency relationship.
In particular, 
the coordinator will only execute a downstream function (e.g., \texttt{defrunAuditRule} in Figure~\ref{fig:finra-workflow}) 
after all its upstream functions (\texttt{fetchPortfolioData} and \texttt{fetchMarketData}) finish. 

\subsection{Fork-aware serverless platform}
\label{sec:fork-aware}

Being aware of {\sys}, 
the platform can leverage parents that have prepared themselves via \texttt{fork\_prepare}
(we term them as \emph{seed}s in this paper) to accelerate function startup and state transfer. 
Besides, it is also responsible for reclaiming the seeds.
Based on the use cases, 
we further categorize seeds into two classes.  
1) For seeds that are used for boosting function startups, 
the frequency of reclamation is low. 
Hence, we name them \emph{long-lived} seeds and 
use a coarse-grained reclamation scheme (\textsection{\ref{sec:serverless-long-seed}}).
2) For seeds that are used for state transfer, 
they only live during the lifecycle of a serverless workflow.
We name them \emph{short-lived} seeds and use 
a fine-grained fork tree-based mechanism 
to free them (\textsection{\ref{sec:fork-tree}}). 

The steps to accelerate {\fn} with {\sys} are:
(1) Extend the {\fn} coordinator to send prepare/resume 
requests to the invoker to fork containers if necessary and 
(2) Instrument FDK so that it can recognize the 
new (fork) requests from the coordinator 
(\eg{line 12--16 in Figure~\ref{fig:finra-fork-code} (b)}). 
Since the extensions to the FDK are trivial, 
we focus on describing the extensions to the coordinator. 

\begin{figure}[!t]
        \centering
        \hspace{-1mm}
        \includegraphics[scale=1.1]{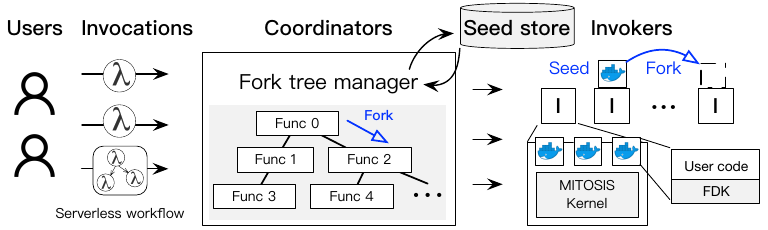} \\[10pt]
        \begin{minipage}{1\linewidth}
        \caption{\small\emph{Integrating {\sys} to {\fn}. 
        The gray boxes are our added (or extended) components. 
        }}
        \label{fig:fn}
        \end{minipage} \\[-15pt]
        \end{figure}
		
\stitle{Fork-aware coordinator.} 
For a single function call, 
the coordinator first looks up an available (long-lived) seed. 
The locations of seeds are stored at a \emph{seed store}.
If one seed is available, 
it sends a fork resume request to the invoker. 
Otherwise,
we fallback to the vanilla function startup mechanism. 

During workflow execution, 
the coordinator dynamically creates short-lived based on state transfer relationship. 
Specifically,
it will tell the invoker to call \texttt{fork\_prepare} 
if it executes an upstream function in the workflow.
The prepared results are piggybacked in the reply of the function.
Afterward, 
the coordinator can use \texttt{fork\_resume} to 
start downstream functions,
which transparently inherit the pre-materialized results of the upstream one. 

Note that one function may have multiple upstream functions 
(\eg{\emph{run AuditRule} in Figure~\ref{fig:finra-workflow}). 
For such cases, 
we require the user to specify which function to fork by annotating the workflow graph
or fuse the upstream functions.  

\subsection{Long-lived seed management}
\label{sec:serverless-long-seed}

\stitle{Deployment.} 
We deploy long-lived seeds as cached containers
because they naturally load the function's working set into the memory. 
If the invoker decides to cache a container, 
it will call \texttt{fork\_prepare} to generate a seed.
Note that 
we must also adjust {\fn}'s cache policy to be fork-aware. 
For example, {\fn} always caches a container if it experiences a coldstart,
which is unnecessary considering {\sys}
because the fork can accelerate startups more resource-efficiently. 
Therefore, we only cache the first container facing coldstart across the platform. 
Moreover, we also detect whether a container is a multi-hop one, 
\ie{forked from a long-lived seed}. 
We don't cache such containers as they are short-lived seeds.

\stitle{Seed store.}
To find the seed information, 
we record a mapping between function name 
and 
the corresponding seed's RDMA address,
the \texttt{handle\_id} and \texttt{key} (the latter two are returned by \texttt{fork\_prepare}) at the coordinator.
We also record the time when the seed was deployed, 
which is necessary to prevent the coordinator forking from a 
near-expired cache instance. 
The seed store can be co-located with the coordinator 
or implemented as a distributed key-value store.

\stitle{Reclamation.}
Similar to Caching, 
the long-lived seeds are reclaimed by timeout.
Unlike Caching, seeds can have a much longer keep-alive time 
(\eg{10 minutes vs. 1 minute})
since they consume orders of magnitude smaller memory.
The coordinators can renew the seed 
if it doesn't live long enough for the forked function.

\subsection{Fork tree and short-lived seed management}
\label{sec:fork-tree}

\noindentstitle{Fork tree granularity and structure.}
Each serverless workflow 
has a dedicated fork tree stored and maintained at the coordinator executing it.
The upper-layer nodes in the tree correspond to the upstream functions (parents) 
in the workflow
and the lower-layer nodes represent the downstream functions (children). 
Each node encodes the container IDs and locations,
which is sufficient for the coordinator to reclaim the corresponding seed. 

\stitle{Fork tree construction and destroy.}
The construction of the fork tree is straightforward: 
After the coordinator forks a new child from a short-lived seed,
it will add the seed to the tree.
When all functions in the tree finish, 
{\sys} will reclaim all the nodes except for the root node: 
the root node can be a long-lived seed and {\sys} will not reclaim it. 

\stitle{Fault tolerance.}
The fork tree should be fault-tolerant to 
prevent memory leakage caused by dangling seeds. 
Replicating the tree with common replication protocols (\eg{Paxos~\cite{lamport2001paxos}) 
can tolerate the failure, 
but adds non-trivial overheads during the workflow execution. 
Observing that serverless functions 
have a maximum lifetime (\eg{15 minutes in AWS Lambda~\cite{aws-lifetime}}), 
we use a simple timeout-based mechanism to tolerate the failures.
Specifically, 
invokers will periodically garbage collect short-lived seeds
if they run beyond the function's maximum allowed runtime.

\subsection{Limitation}
First, fork still needs a long-lived seed to quickly bootstrap others. 
If no seed is available, 
we can leverage existing approaches that optimize coldstart (\eg{FaasNET~\cite{DBLP:conf/usenix/WangCTWYLDC21}}) 
to first start one. 
Second, fork only enables a read-only state transfer. 
Yet, it is sufficient for serverless workflow---the dominant function composition method. 
Finally, fork cannot transfer states between multiple upstream functions. 
Thus, {\sys} must fuse these upstream functions into one or fallback to messaging 
(see \texttt{Portfolio} in Figure~\ref{fig:finra-fork-code}) for such cases.
We are addressing this limitation by further introducing a \emph{remote merge} primitive 
to complement the remote fork.

\begin{figure*}[!t]
    \hspace{-1mm}
    \includegraphics[scale=1.08,left]{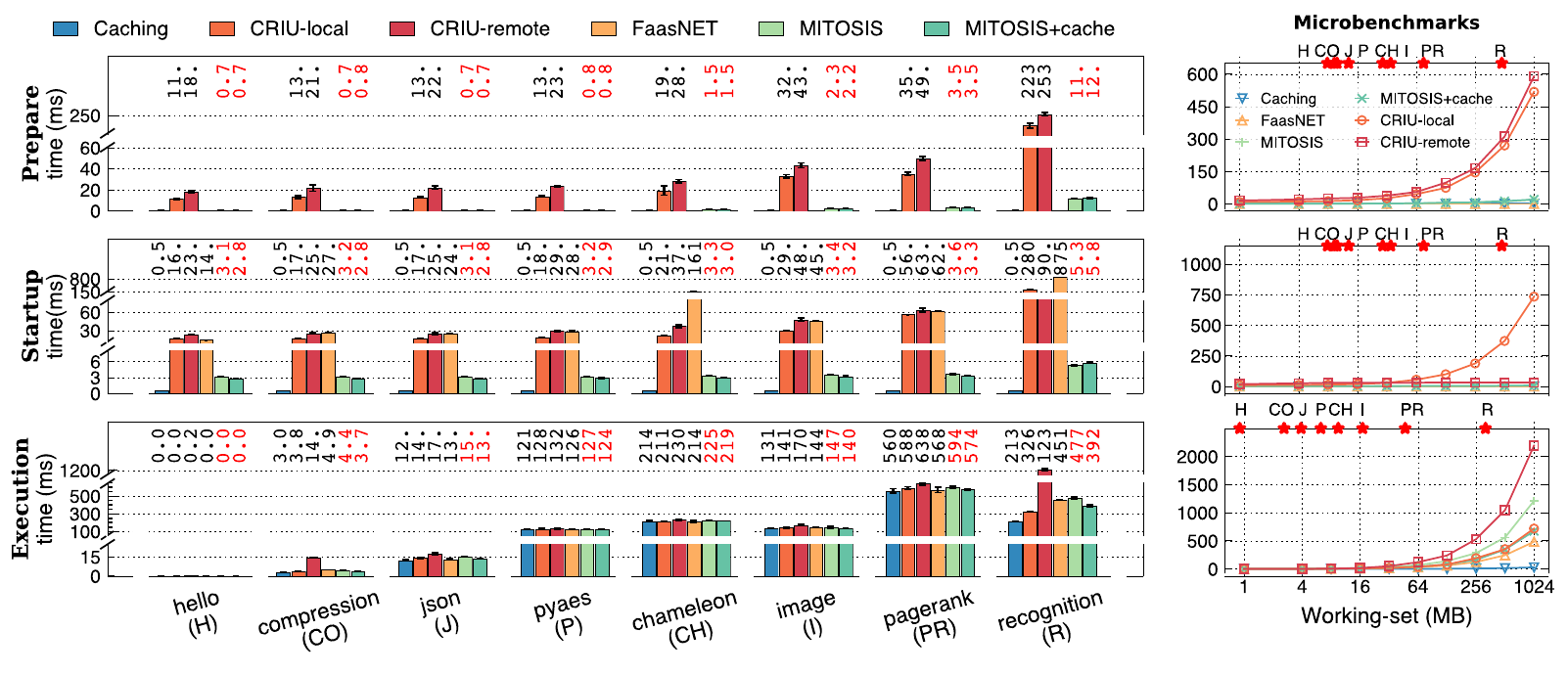} \\[3pt]
             \begin{minipage}{1\linewidth}
                 \caption{\small\emph{(a) End-to-end latency comparisons of {\sys} and baselines. 
                 (b) Analyses of different phases using microbenchmarks. 
                 Note that the working set of the execution is smaller than the 
                 prepare and startup because child only touches a subset of the 
                 parent's memory. 
                 }}
     \label{fig:eval-e2e}
     \end{minipage} \\[-10pt]
\end{figure*}

\section{Evaluation} 
\label{sec:eval}
\label{sec:eval-setup}

\noindentstitle{Experimental setup.} 
We conduct all our experiments on a local cluster with 24 machines.
Each machine has two 12-core Intel Xeon E5-2650 v4 processors and 128GB of DRAM. 
16 machines are connected to two Mellanox SB7890 100Gbps switches 
with two 100\,Gbps ConnectX-4 MCX455A InfiniBand RNICs. 
We use them as invokers to execute the serverless functions. 
Nodes without RDMA are left as coordinators.

\stitle{Comparing targets.} 
The evaluating setups of {\sys} and its baselines 
are listed as follows. 
Note that we apply our generalized lean container (\textsection{\ref{sec:resume}}) 
to all the systems to hide the cost of containerization. 
\begin{enumerate}[leftmargin=*]
    \itemsep0.5em    
    \setlength{\itemindent}{0em} 
    \item \textbf{\emph{Caching}} is the de facto warmstart technique that provides
    a near-optimal function startup.

    \item \textbf{\emph{CRIU-local}} leverages CRIU~\cite{criu} to implement remote fork
	(see Figure~\ref{fig:overview} (a))
	and stores all files in an in-memory local filesystem (tmpfs). 
	The file is transferred via our optimized transfer library with one-sided RDMA. 
    We also apply existing on-demand restore optimization~\cite{DBLP:conf/eurosys/WangHW19}.

    \item \textbf{\emph{CRIU-remote}} leverages CRIU and a distributed file system 
	for the remote fork (see Figure~\ref{fig:overview} (b)). 
    We use Ceph~\cite{ceph-rdma}---a state-of-the-art production DFS that embraces RDMA. 
    We also apply optimizations from CRIU-local: in-memory storage and on-demand restore. 

    \item \textbf{\emph{FaasNET}}~\cite{DBLP:conf/usenix/WangCTWYLDC21}
	optimizes the container image pulling of coldstart with function trees. 
    We evaluate an optimal setup of FaasNET (for performance) that pre-provisions the images 
	at all the invokers.\footnote{\footnotesize{Confirmed by the authors.}}

    \item \textbf{\emph{\sys}} 
	is configured with on-demand execution and reads all pages from remote
    with a prefetch size of one.

    \item \textbf{\emph{{\sys}+cache}}
    is the version of {\sys} that always caches and shares the fetched pages among children.
    It essentially fallbacks to the local fork. 
\end{enumerate}

\vspace{-2mm}
\stitle{Functions evaluated.}
We chose functions from representative serverless benchmarks
(i.e., ServerlessBench~\cite{DBLP:conf/cloud/YuLDXZLYQ020}, 
FunctionBench~\cite{DBLP:conf/cloud/KimL19}, 
and SeBS~\cite{DBLP:conf/middleware/CopikKBPH21}),
which cover a wide range of scenarios, 
including simple function (\emph{hello}/{H}---print `Hello world'), 
file processing (\emph{compression}/{CO}---compress a file), 
web requests (\emph{json}/{J}---(de)serialize json data,  
\emph{pyaes}/{P}---encrypt messages, \emph{chameleon}/{CH}---generate HTML pages), 
image processing (\emph{image}/{I}---apply image processing algorithms to an image), 
graph processing (\emph{pagerank}/{PR}---execute the pagerank algorithm on a graph) 
and machine learning (\emph{recognition}/{R}---image recognition using ResNet). 
These functions are written in python---the dominant serverless language~\cite{report-state-of-serverless}.
Besides, we also use a synthetic \emph{micro-function} that touches a variant portion 
of the memory to analyze the overhead introduced by {\sys}. 
It is written in C to minimize the language runtime overhead interference.

\subsection{End-to-end latency and memory consumption} 
\label{sec:eval-e2e}

Figure~\ref{fig:eval-e2e} shows the results of end-to-end latency: 
the left subfigure is the time of different phases of the functions during remote fork,
and the right is each phase's result on micro-function. 
The function request is sent by a single client.
To rule out the impact of disk accesses, we put all the function's related files
(\eg{images used by \emph{image}/I}) in tmpfs. 

\stitle{Prepare time.}
The prepare time is the time 
for the parent to prepare a remote fork. 
For CRIU-local and CRIU-remote, it is the time to checkpoint a container. 
For variants of {\sys}, it is the \texttt{fork\_prepare} time. 
Caching and FaasNET do not have this phase because they do not support fork. 

{\sys} is orders of magnitude faster in preparation 
than CRIU-local and CRIU-remote. 
On average, it reduces the prepare time by 94\%. 
For example, {\sys} prepared a 467\,MB \emph{recognition}/R container in 11\,ms, 
while CRIU-local and CRIU-remote took 223\,ms and 253\,ms, respectively. 
The variants of CRIU are bottlenecked by copying the container state
from the memory to the filesystems. 

\begin{figure*}[!t]
    \hspace{-1mm}
    \includegraphics[scale=1.12,center]{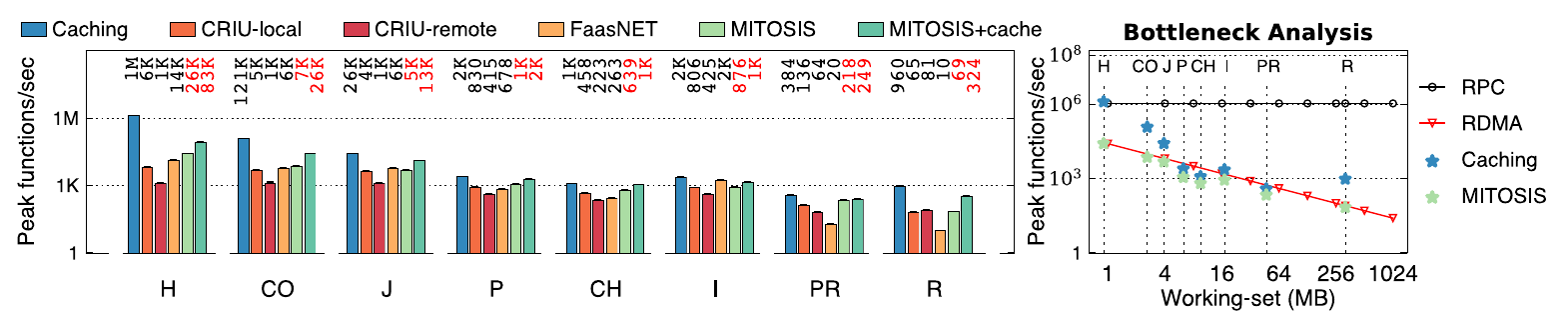} \\[-4pt]
             \begin{minipage}{1\linewidth}
                 \caption{\small\emph{(a) Peak throughput comparisons of {\sys} and baselines.
                 (b) Bottleneck analysis of {\sys} using a single parent seed.
                 }}
     \label{fig:eval-e2e-thpt}
     \end{minipage} \\[-10pt]
\end{figure*}

\stitle{Startup time.} 
We measure the startup time as the time between 
an invoker receiving the function request 
and the time the first line of the function executes. 
As shown in the middle of Figure~\ref{fig:eval-e2e}, 
caching is the fastest (0.5\,ms) 
because starting a cached container only requires a simple unpause operation. 
{\sys} comes next, it can start all the functions within 6\,ms. 
It is up to 99\%, 94\%, and 97\% (from 98\%, 86\%, and 77\%) faster than 
CRIU-local, CRIU-remote, and FaasNET, respectively. 
The startup time of {\sys} is dominated by the generalized lean container setup time 
since reading the descriptor with RDMA is extremely fast with our fast descriptor fetch protocol. 

The startup of CRIU-local is dominated by copying the entire file (shown in Figure~\ref{fig:eval-e2e} (b)). 
Using CRIU-remote avoids transferring the file,
but the overhead of communicating with the DFS meta server (from 23--90\,ms) is still non-trivial. 
Compared to CRIU-remote, 
{\sys} can directly read the container metadata (descriptor) from the remote machine's kernel. 
Finally, the startup cost of FaasNET (coldstart) is dominated by the runtime initialization of the function, 
as we skipped the image pull process of it. 
The overhead depends on the application characteristics. 
For example, \emph{recognition}/R requires loading a ResNet model from PyTorch, 
which takes 875\,ms.
Other techniques can skip the loading process
since the model has been loaded in the parents or the cached containers.

Note that the results of CRIU-remote and FaasNET 
are not significantly higher in the startup microbenchmark (Figure~\ref{fig:eval-e2e} (b)).
For CRIU-remote, it is because the time (40ms) is relatively small compared to CRIU-local 
(\textgreater191ms for working-set larger than 256MB). 
For FaasNET, we use native language in the microbenchmark,
so it doesn't suffer from the runtime initialization and library loading costs 
of the application functions in Figure~\ref{fig:eval-e2e} (a).  

\stitle{Execution time.}
For function execution, 
{\sys} is up to 2.24$\times$, 1.46$\times$ and 1.14$\times$
(from 1.04$\times$, 1.04$\times$, and 1.02$\times$) slower
than Caching, CRIU-local and FaasNET, respectively, except for \emph{hello}/H.    
The overhead is mainly due to page faults and reading remote memory,
which is proportional to the function working set (see Figure~\ref{fig:eval-e2e} (b)). 
Consequently, the overhead is most significant in \emph{recognition}/R 
that reads 321\,MB of the parent memory:
{\sys} is 2.24$\times$ (477 vs. 213\,ms) and 1.46$\times$ (477 vs. 326\,ms)
slower than Caching and CRIU-local, respectively.
CRIU-local is faster since it reads files from the local memory (tmpfs).
To remedy this, {\sys}+cache reduces the number of remote memory accesses 
by reading from the local cached copies of the remote pages. 
It improves performance by up to 17\%, 
making {\sys} close to or better than CRIU-local and FaasNET during execution. 
Note that Caching is always optimal (\ie{faster than FaasNET and CRIU-local}) because 
it has no page fault overhead. 
Finally, {\sys} is up to 3.02$\times$ (from 1.02$\times$) faster than CRIU-remote 
thanks to bypassing DFS for reading remote pages.

\begin{figure}[!t]
    \hspace{-2.1mm}
    \begin{minipage}{.48\linewidth}
            \includegraphics[scale=1.24,left]{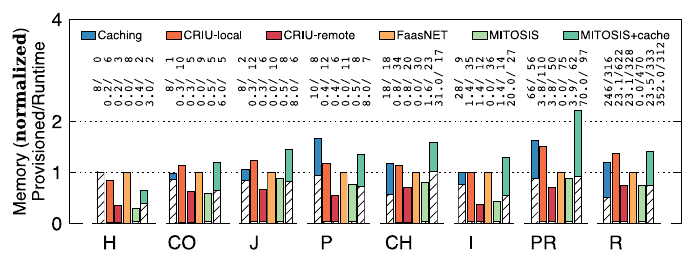} 
            \end{minipage} 

            \begin{minipage}{1\linewidth}
                \caption{\small\emph{The per-function memory usage (in MB) for each technique
                before running (hatched) and during runtime (colored).
                }}
    \label{fig:eval-e2e-mem}
    \end{minipage} \\[-5pt]
    \end{figure} 

\stitle{Memory consumption.} 
Figure~\ref{fig:eval-e2e-mem} reports the amortized per-machine memory consumed 
for each function categorized by provisioned memory (before running) and runtime memory. 
An ideal serverless platform should use minimal provisioned memory for each function.
On average, 
{\sys} only consumes 6.5\% of the provisioned memory (one cached instance across 16 machines)
while Caching requires at least 16 instances.
CRIU-local/remote consumes a slightly lower memory (77\% on average) 
than {\sys}, 
because it reuses the local OS's shared libraries to prevent storing them 
in the checkpointed files. 
This works at the cost of requiring storing all the function's required libraries on all the machines,
otherwise the restored container will fail. 
For the same reason, {\sys} consumes a slightly larger runtime memory (8\% on average)
than CRIU-remote. 
Yet, its runtime memory is smaller than CRIU-local 
because the CRIU-local will read the entire file before it can execute the function.


\subsection{Bottleneck analysis and throughput comparisons}

\noindentstitle{Bottleneck analysis.}
Using a single seed function is ideal for resource usage. 
However, the parent-side network bandwidth (RDMA) 
and two RPC threads can become the bottleneck.
Meanwhile, 
{\sys} is also bottlenecked by the aggregated client-side CPU resources 
processing the function logic. 
The peak client-side performance for each function 
is the peak throughput of running functions with Caching. 

Figure~\ref{fig:eval-e2e-thpt} (b) analyzes the impact 
of the above factors. 
We utilize all 16 invokers to achieve the peak throughput. 
For H, CO, J, and R, 
RDMA is the bottleneck.
For example, \emph{recognition}/R touches
321\,MB of the parent's memory, 
so the RDMA (200\,Gbps) can only serve (ideal) 80 forks/sec. 
Thus, {\sys} achieves 69 reqs/sec and is lower than Caching (960 reqs/sec). 
In contrast, 
if the children CPU is the bottleneck, 
{\sys} is similar to Caching (P, CH, I, and PR). 
For example, 
Caching can only execute 384 reqs/sec for \emph{pagerank}/PR.
In comparison, 
RDMA can handle an ideal 544 PR forks/sec (the working set is 47\,MB). 
Thus, {\sys} can achieve a slightly lower throughput (249 reqs/sec).
Finally, the RPC would never become the bottleneck:
two kernel threads can handle up to 1.1 million reqs/sec,
which is always faster than RDMA for working set from 1\,MB to 1\,GB. 


\stitle{Throughput comparison.}
Figure~\ref{fig:eval-e2e-thpt} (a) further compares the peak throughput 
of different approaches. 
Note that we exclude the prepare phase of CRIU---otherwise,
it will be bottlenecked by this phase.
{\sys} is up to 8.0$\times$ (from 2.1$\times$) faster than CRIU-local,
thanks to avoiding the whole file during the restore phase. 
Compared with CRIU-remote, 
{\sys} is also up to 20.4$\times$ (from 2.1$\times$) faster except for R (69 vs. 81):
CRIU-remote reads a smaller amount of remote memory because 
it reuses local copies of the shared libraries.
R has the largest working set, so it is mostly affected by the network.
For the others, {\sys} is faster as it bypasses the overhead of DFS.
We omit the comparison between {\sys} and Caching, 
which has been studied in the bottleneck analysis.

\begin{figure}[!t]
    \hspace{-7pt}
    \includegraphics[scale=1.25]{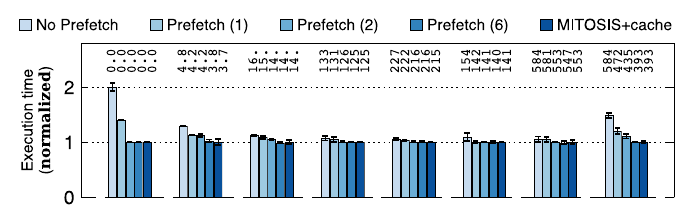} \\[0pt]
    \includegraphics[scale=1.25]{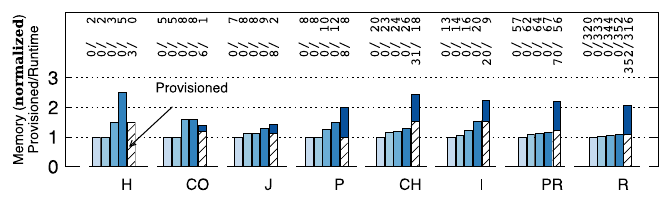} \\[5pt]
    \begin{minipage}{1\linewidth}
    \caption{\small\emph{Effects of the number of pages prefetched per-fault on 
    (a) execution time (in ms) and (b) memory consumption (in MB).}}
    \label{fig:eval-prefetch}
    \end{minipage} \\[-5pt]
    \end{figure} 

\subsection{Effects of prefetching}
\label{sec:eval-prefetch}

We next explore how the prefetch number affects {\sys} in Figure~\ref{fig:eval-prefetch} (a).
As we can see, prefetching can significantly improve the 
execution time of functions:
prefetching 1, 2, and 6 pages improve the average time 
by 10\%, 16\%, and 18\% (up to 30\%, 50\%, and 50\%), respectively. 
More importantly, 
a small prefetch size (6) can achieve a near-identical performance 
as the optimal, \ie{no remote access, ({\sys}+cache)}. 
Note that for small prefetch size the cost 
to the throughput is negligible, so we omit the results.

Prefetching has additional runtime memory consumption: 
as shown in Figure~\ref{fig:eval-prefetch} (b), 
prefetching 1, 2, and 6 consumes average
1.1$\times$, 1.3$\times$, and 1.5$\times$ 
(up to 1.15$\times$, 1.6$\times$, and 2.5$\times$) more memory 
than no prefetching. 
Therefore, we currently adopt a prefetch size of 1 to reduce runtime memory usage.

\begin{figure}[!t]
        \centering
        \includegraphics[scale=0.95]{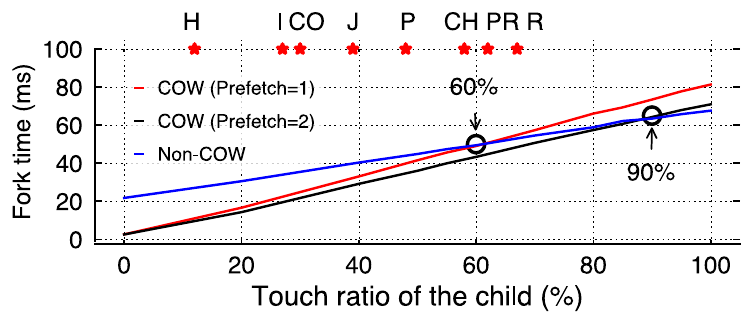} 
        \centering
        \includegraphics[scale=1]{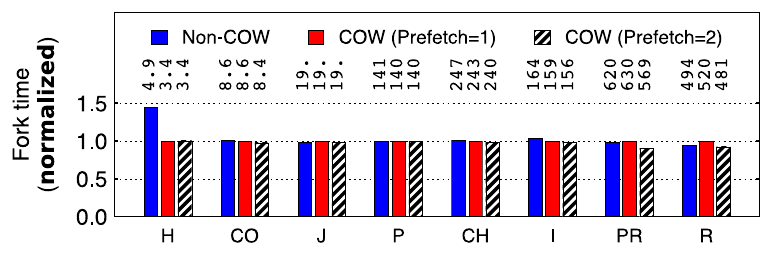} \\[5pt]
        \begin{minipage}{1\linewidth}
        \caption{\small\emph{Effects of COW to latencies on (a) the micro-function 
		(with a 64\,MB parent working set) and (b) serverless functions.        
        }}
        \label{fig:eval-cow}
        \end{minipage} \\[-5pt]
        \end{figure} 

\subsection{Effects of copy-on-write (COW)}
\label{sec:eval-cow}        
        
{\sys} reads the child's pages in an on-demand way (copy-on-write).
This section presents the benefits and costs of COW compared to 
a non-COW design---the child will read all the parent's memory before 
executing the functions. 

\stitle{Latency.}         
Figure~\ref{fig:eval-cow} reports the latency results.  
The benefit of COW in latency depends on the amount 
of the parent's memory touched by the child (touch ratio):
the cross points in the microbenchmark are 60\% and 90\%
when the prefetch size is 1 and 2, respectively. 
For larger prefetch size, the cross point is close to 100\%. 
Non-COW has a longer startup time due to extra remote memory reading, 
but it is more efficient in reading pages with RDMA because
it can batch multiple paging requests~\cite{DBLP:conf/usenix/KaliaKA16}. 
Nevertheless, serverless functions typically have a moderate touch ratio (\ie{$<$ 67\%}).
Therefore, 
COW has averages of 8.7\% (from 0.6\% to 44\%) and 3.7\% (from -5\% to 31\%) 
lower latency than Non-COW when the prefetch size is 1 and 2, respectively. 

\begin{figure}[!t]
        \centering
        \includegraphics[scale=0.95]{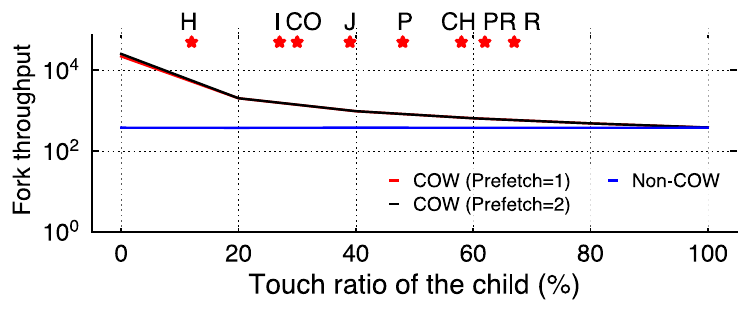} 
        \centering
        \includegraphics[scale=1]{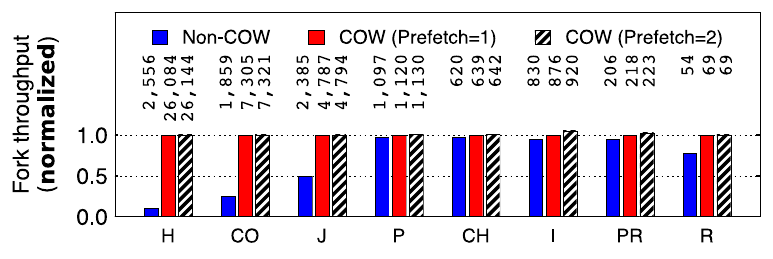} \\[5pt]        
        \begin{minipage}{1\linewidth}
        \caption{\small\emph{Effects of COW to peak thpt on (a) the micro-function 
		(with a 64\,MB parent working set) and (b) serverless functions.}}
        \label{fig:eval-cow-thpt}
        \end{minipage} \\[-0pt]
        \end{figure} 

\stitle{Throughput.} 
Figure~\ref{fig:eval-cow-thpt} further reports the throughput results. 
Unlike latency, 
COW is always faster in throughput (except for 100\% touch ratio) 
because non-COW will issue more RDMA requests.  
Consequently, 
COW is 1.03X--10.2X faster than Non-COW on serverless functions. 
      
\begin{figure}[!t]
            \includegraphics[left, scale=1.1]{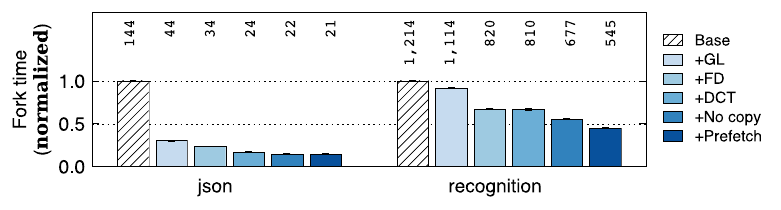} \\[-2pt]
            \begin{minipage}{1\linewidth}
            \caption{\small\emph{Effects of optimizations applied by {\sys}.}}
            \label{fig:e2e-factor-analysis}
            \end{minipage} \\[-20pt]
            \end{figure}

\begin{figure*}[!t]
    \hspace{-1mm}
    \includegraphics[scale=1.25,center]{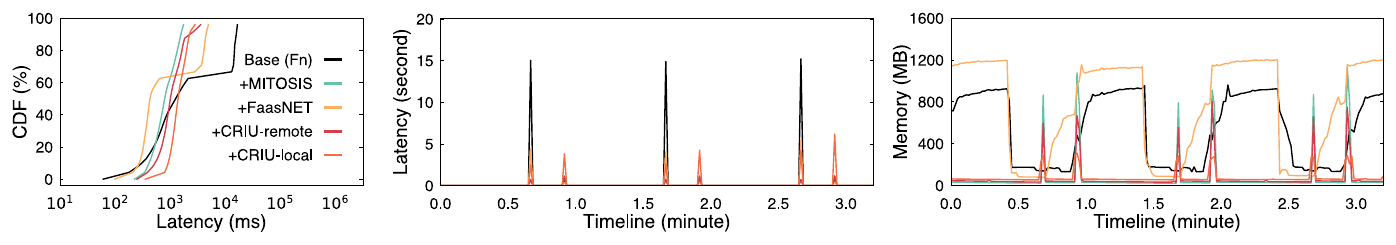} \\[-4pt]
             \begin{minipage}{1\linewidth}
                 \caption{\small\emph{(a) The latency CDF, (b) average latency, 
                 and (c) memory consumption timelines on image processing \textup{(I)} under spikes.}}
                 \label{fig:eval-loadspike}
     \end{minipage} \\[-12pt]
\end{figure*}          

\begin{figure}[t]
    \includegraphics[scale=1.0,center]{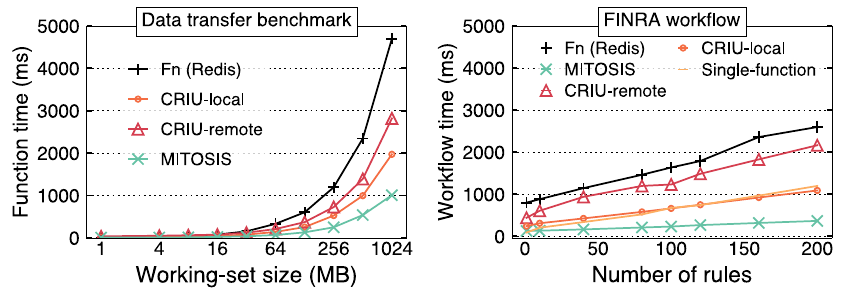} \\[5pt]
    \begin{minipage}{1\linewidth}
    \caption{\small{\emph{(a) The state-transfer performance between two functions 
    and (b) performance of FINRA. }}}
    \label{fig:eval-state}
    \end{minipage} \\[-15pt]
    \end{figure}             

\subsection{Effects of optimizations}
\label{sec:eval-other}
Due to space limitation,
Figure~\ref{fig:e2e-factor-analysis} briefly shows the effects of optimizations 
introduced in \textsection{\ref{sec:design}}
on the end-to-end fork time using a short function (\emph{json}/J) and 
a long function (\emph{recognition}/R).
First, generalized lean container (+GL)
reduced a fixed offset of the latency (100\,ms) to all the functions
compared with a baseline of using runC~\cite{runc}. 
Compared with RPC, 
fast descriptor fetch with one-sided RDMA (+FD)
further contributes 10\% and 25\% latency reduction for both functions.
The improvement is more obvious for R because its descriptor is much larger (1.3\,MB vs. 31\,KB).
Using DCT instead of RC reduced a 10--20\,ms to the functions, 
and directly exposing the physical memory with RDMA instead of copying them (+no copy)
further reduced the fork time by 12\% and 20\% for J and R, respectively. 
Finally, prefetching (+prefetch) shortens the time by 9\% and 15\%.    

\subsection{State-transfer performance}

\noindentstitle{Microbenchmark.}
We use the data-transfer testcase (5) in ServerlessBench~\cite{DBLP:conf/cloud/YuLDXZLYQ020}
to compare different approaches to transfer states between two remote functions.
As shown in Figure~\ref{fig:eval-state} (a), {\sys} is up to 1.4--5$\times$ faster than Fn, 
which leverages Redis to transfer data between functions,
when transferring 1\,MB--1\,GB data.
Note that we exclude the data (de)serialization overhead (by skipping the phase) 
and coldstart overhead (by pre-warming the containers) in Fn.
Otherwise, the gap between Fn and {\sys} would become larger. 
Compared to CRIU-local/remote, {\sys} is faster
thanks to the design for a fast remote fork (see \textsection{\ref{sec:eval-e2e}}).

\stitle{Application: FINRA.} 
We next present the performance of {\sys} on FINRA~\cite{finra},
whose workflow graph is shown in Figure~\ref{fig:finra-workflow}.
We manually fuse the \texttt{fetchPortfolioData}
and \texttt{fetchMarketData} into one function to fully leverage 
remote fork for {\sys} and CRIU variants.  
For {\fn}, functions use Redis to transfer states. 
Figure~\ref{fig:eval-state} (b) reports the end-to-end 
latency w.r.t the number of instances of \texttt{runAuditRule},
where FINRA spawns about 200 instances~\cite{finra-aws}. 
We select the market data from seven stocks, resulting in a total 6\,MB 
states transferred between functions.

As we can see, 
{\sys} is 84--86\%, 47--66\% and 71--83\% faster 
than the baseline Fn, CRIU-local and CRIU-remote, respectively. 
Note that we have pre-warmed Fn to prevent the effects of 
coldstart---which is unnecessary for {\sys}. 
Fn is bottlenecked by Redis (27\,ms) and
data serialization and de-serialization (600\,ms). 
{\sys} has no such overhead and it further makes state
transfer between machines optimal via RDMA. 
Moreover, {\sys} can scale to a distributed setting 
with little COST~\cite{DBLP:conf/hotos/McSherryIM15}---it 
can outperform a single-function sequentially processing all the rules (Single-function).
This is because {\sys} can concurrently run functions across machines 
with minimal cost transferring data between them. 

\subsection{Performance under load spikes}
\label{sec:load-spike}

Finally, we evaluate the performance of {\sys} under load spike using \emph{image}/I 
on the real-world traces (\texttt{660323}~\cite{DBLP:conf/usenix/ShahradFGCBCLTR20}).
Figure~\ref{fig:eval-loadspike} (a) summarizes the latency CDFs. 
The 99$^{th}$ percentile latency of {\fn}+{\sys} is
73.64\% and 89.08\% smaller than {\fn}+FaasNET and {\fn}, respectively, 
thanks to avoiding the coldstart with remote fork. 
Nevertheless, its median latency is 1.85$\times$ longer than FaasNET (799\,ms vs. 430\,ms),
because FaasNET leverages Caching and has a 65.1\% cache hit during spikes.
However, Caching incurs non-trivial memory consumptions:
Fn (and Fn+FaasNET) will cache a container for 30 seconds if it is a coldstart,
resulting in a significant amount of memory usage (see Figure~\ref{fig:eval-loadspike} (c)).
In comparison, {\sys} only caches a single seed
and saves orders of magnitude memory during the idle time.
For example, at time 2.3\,min, 
{\sys} only consumes 29\,MB memory per-machine, 
which is 3\% and 2\% of Fn (914\,MB) and Fn+FaasNET (1,199\,MB), respectively. 

\section{Discussion}
\label{sec:dislim}

\noindentstitle{Seed placement and selection policies.}
We currently choose a random placement policy.
A better policy may further consider 
network topology and system-wide load balance. 
Meanwhile, we simply choose the first container experiencing coldstart 
as the long-lived seed,
yet, a better selection policy should further consider the status of the running container. 
For instance, recent works have discovered that containers may need multiple invocations to warm up properly~\cite{DBLP:conf/hotos/CarreiraKBF21, DBLP:conf/eurosys/ShinKM22}, 
\eg{to JIT a function written in a managed language}. 
Therefore, choosing a properly warm-up container as the seed
can significantly improve the function performance after the fork. 
As these policies are orthogonal to {\sys}, we plan to investigate them in the future.

\stitle{Frequency and cost of the fallback.}
The frequency of the fallback can significantly impact the performance 
of remote fork. 
During our experiment, 
we encountered no fallback because the parent (cached container) 
must have loaded all the children's memory. 
Nevertheless, fallback can happen in corner cases (\eg{swap}). 
The per-page overhead is 22$\times$ (65 vs. 3\,$\mu$s) due to the cost of 
RPC and loading the page from the disk (SSD). 
Currently, one fallback handler can process 16K paging requests per second,
so it will not become the bottleneck. 

\stitle{The benefits of implementing {\sys} in the kernel. }
We choose to implement {\sys} in the kernel for performance considerations. 
First, a user-space solution cannot directly access the physical memory of the container, 
so it pays the checkpointing overhead (see \textsection{\ref{sec:rfork}}). 
Moreover, the kernel can establish RDMA connections more efficiently (see KRCore~\cite{krcore}),
and the kernel-space page fault handler is much faster than the user-space fault handler.

\section{Related Work}
\label{sec:related}

\noindentstitle{Optimizing serverless computing.}
{\sys} continues the line of research on optimizing serverless computing,
including but not limited to accelerating function startups~\cite{sock,sand,FAASM,du2020catalyzer,DBLP:conf/asplos/Ustiugov0KBG21,DBLP:conf/eurosys/SaxenaJSKA22,DBLP:conf/usenix/WangCTWYLDC21},
state transfer~\cite{cloudburst, DBLP:journals/usenix-login/KlimovicWSTPK19,DBLP:conf/nsdi/PuVS19,faastlane,sand, DBLP:conf/usenix/MahgoubSMKCB21},
stateful serverless functions~\cite{Beldi,DBLP:conf/sosp/JiaW21},
transactions~\cite{10.1145/3464298.3493392},
improving the cost-efficiency~\cite{Harvested-severless,DBLP:conf/asplos/FuerstS21,DBLP:conf/eurosys/RzadcaFSZBKNSWH20, DBLP:journals/cal/LiLMITKK21, DBLP:conf/osdi/QiuBJKI20,DBLP:conf/apsys/FinglerAR19,DBLP:conf/cloud/DukicBSA20},
and others~\cite{DBLP:conf/eurosys/SreekantiWCGHF20, DBLP:conf/cloud/ZhangFPS20, 
DBLP:conf/cloud/KaffesYK19,DBLP:conf/asplos/DuLJXZC22,DBLP:conf/asplos/JiaW21,DBLP:conf/nsdi/AgacheBILNPP20,DBLP:conf/osdi/ThorpeQETHJWVNK21,
DBLP:conf/eurosys/LyuCAP022, DBLP:conf/asplos/YangZLZLZCL22, DBLP:conf/sc/ZhaoYLZL21}.
Most of these works are orthogonal to {\sys}.
Nevertheless, we believe they can also benefit from our work. 
In particular, we propose to use the remote fork abstraction to 
simultaneously accelerate function startups and state transfer,
which is critical to all serverless applications. 
For our closest related works, 
we have also extensively compared them in \textsection{\ref{sec:bg}}. 

Though the implementation of Linux fork may not be optimal
in some scenarios~\cite{snapshot-fuzzing,fork-road,fork-on-demand},
it has been shown to be suitable for serverless functions~\cite{sand,du2020catalyzer}.
Therefore, we still choose to generalize the fork abstraction 
to accelerate functions running across machines. 

\stitle{Checkpoint and restore (C/R).}
C/R has been investigated by {OS}es for a long time~\cite{egwutuoha2013survey,litzkow1997checkpoint}.
e.g., KeyKOS~\cite{keykos}, EROS~\cite{eros}, Aurora~\cite{Aurora}
and others~\cite{hargrove2006berkeley,laadan2010linux,criu,zhong2001crak,vas-criu,DBLP:conf/sigmod/ArmenatzoglouBB22,DBLP:journals/corr/abs-2105-14845,https://doi.org/10.48550/arxiv.2206.13444}.
Aurora~\cite{Aurora} leverages C/R to realizing efficient single level store, 
it introduces techniques including system shadowing for efficient incremental checkpointing. 
{\sys} eliminates checkpointing in the context of remote fork via OS-RDMA co-design.
VAS-CRIU~\cite{vas-criu} also noticed the overhead of C/R introduced by filesystems. 
It leverages multiple independent address spaces (MVAS)~\cite{DBLP:conf/asplos/HajjMZMAFHRS16} 
to bypass the filesystem for C/R on a single machine. 
We further use kernel-space RDMA to build a global distributed address space 
and scale fast C/R to a distributed setting. 

\stitle{Remote fork (migrations). }
Besides using C/R for remote fork~\cite{smith1987implementing,criu-rfork}, 
{\sys} is also inspired by works on 
virtual machine fork (SnowFlock~\cite{DBLP:conf/eurosys/Lagar-CavillaWSPRLBS09})
and migrations~\cite{DBLP:conf/hpdc/Al-KiswanySSR11, DBLP:conf/nsdi/ClarkFHHJLPW05,DBLP:conf/dsn/GuHXCZGL17,DBLP:conf/vee/HinesG09,10.1145/1618525.1618528, DBLP:journals/csur/MilojicicDPWZ00, litzkow1992supporting}, 
just to name a few.
For example, the {\sys} container descriptor is inspired by the VM descriptor used in SnowFlock,
which only captures the critical metadata used for instantiating a child container at the remote side.
We further consider the opportunities and challenges when embracing RDMA and serverless computing.
Therefore, we believe our techniques can benefit existing works not using RDMA.

\stitle{RDMA-based remote paging and RDMA multicast.}
Reading pages from remote hosts via RDMA is a not so new technique in modern OSes~\cite{DBLP:conf/eurosys/AmaroBLOAPRS20,DBLP:conf/nsdi/GuLZCS17,DBLP:conf/usenix/AguileraACDGNRS18, DBLP:conf/usenix/MarufC20, DBLP:conf/osdi/ShanHCZ18}. 
For example, Infiniswap~\cite{DBLP:conf/nsdi/GuLZCS17} leverages RDMA 
to build a fast swap device for memory disaggregation. 
Remote regions~\cite{DBLP:conf/usenix/AguileraACDGNRS18} proposes a remote file-like 
abstraction to simplify exposing application's memory with RDMA.
{\sys} further builds efficient remote fork by reading remote pages in a ``copy-on-write'' 
fashion with RDMA.

{\sys} exhibits a pull-based RDMA multicast communication pattern---\eg{multiple 
children pulling from the same parent's memory during load spikes}. 
Push-based RDMA multicast has been extensively studied in the literature~\cite{rdmc,Derecho,DBLP:journals/corr/abs-2110-00886}.
For example,
RDMC~\cite{rdmc} proposes a binomial pipeline protocol 
where a sender can efficiently push data to a group of nodes using RDMA. 
We believe {\sys} can further benefit from researches on pull-based RDMA multicast.

\section{Conclusion}
\label{sec:concl}

We present {\sys}, a new OS primitive for fast remote fork 
by co-designing with RDMA.
{\sys} has two key attributes for serverless computing. 
(1) Startup efficiency: {\sys} is orders of magnitude faster than coldstart
while consuming orders of magnitude smaller resource provisioned compared to warmstart 
(with a comparable performance). 
(2) State transfer efficiency: functions can directly 
access the pre-materialized states from the forked function. 
Extensive evaluations using real-world serverless applications confirmed 
the efficacy and efficiency of {\sys} on commodity RDMA-capable clusters. 
Though we focus on serverless computing in this paper,
we believe {\sys} also shines with other tasks, 
\eg{container migrations}.

\section*{Acknowledgment}
We sincerely thank our shepherd Christopher Rossbach and the anonymous reviewers, 
whose reviews, feedbacks, and suggestions largely strengthen our work. 
We also thank Wentai Li, Qingyuan Liu, 
Zhiyuan Dong, Dong Du, Nian Liu, Sijie Shen, and Xiating Xie for their valuable feedback. 
This work was supported in part by 
the National Key Research \& Development Program of China (No. 2020YFB2104100), 
the National Natural Science Foundation of China (No. 61925206),
Shanghai AI Laboratory, and a research grants from Huawei Technologies. 
Corresponding author: Rong Chen (\burl{rongchen@sjtu.edu.cn}).

\balance

\small{
\bibliographystyle{acm}
\bibliography{rum}
}

\clearpage

\end{document}